\documentclass{emulateapj}
\usepackage{amsmath}

\shorttitle{BAT Catalog and Intrinsic Distributions}
\shortauthors{Butler et al.}

\def\gtrsim{\mathrel{\hbox{\rlap{\hbox{\lower4pt\hbox{$\sim$}}}\hbox{$>$}}}}
\def\lessim{\mathrel{\hbox{\rlap{\hbox{\lower4pt\hbox{$\sim$}}}\hbox{$<$}}}}

\newcommand\swift{{\it Swift}}

\slugcomment{Submitted to ApJ}

\begin{document}

\setlength{\pdfpageheight}{\paperheight}
\setlength{\pdfpagewidth}{\paperwidth}

\title{The Cosmic Rate, Luminosity Function and Intrinsic Correlations of Long GRBs}

\author{Nathaniel R. Butler\altaffilmark{1},
Joshua S. Bloom\altaffilmark{2,3},
\& Dovi Poznanski\altaffilmark{2,4}}
\altaffiltext{1}{Einstein Fellow, Astronomy Department,
University of California, Berkeley, CA, 94720-7450, USA}
\altaffiltext{2}{Astronomy Department,
University of California, Berkeley, CA, 94720-7450, USA}
\altaffiltext{3}{Sloan Research Fellow}
\altaffiltext{4}{Lawrence Berkeley National Laboratory, 1 Cyclotron Road, Berkeley, CA 94720, USA}

\begin{abstract}
We calculate durations and spectral parameters for 207 Swift 
bursts detected by the BAT instrument from April 2007 to August 2009,
including 67 events with measured redshifts.  This is the first 
supplement to our
catalog of 425 Swift GRBs (147 with redshifts) starting from GRB~041220.
This complete and extensive data set, analyzed with a  
unified methodology, allows us to conduct an accurate census of intrinsic GRB energetics, hardnesses, durations, and redshifts.
The GRB world model we derive reproduces well the observables from
both Swift and pre-Swift satellites.
Comparing to the cosmic star formation rate, we estimate that only about 0.1\% of massive stars explode as bright GRBs.
There is strong evidence for evolution in the Swift population at intermediate and high-z, and we can
rule out (at the 5-sigma level) that this is due to evolution in the luminosity function of GRBs.
Instead, the Swift sample suggests a modest propensity for low-metallicity, evidenced by an increase in the rate density with redshift.
Treating the multivariate data and selection effects rigorously, we find a real, intrinsic correlation between $E_{\rm iso}$ and $E_{\rm pk}$ (and possibly also $T_{r45,z}$); 
however, the correlation {\it is not} a narrow log-log relation and its observed appearance 
is strongly detector-dependent.
We also estimate the high-z rate ($3-9$\% of GRBs at z beyond 5) and discuss the extent of a large missing population of low-$E_{\rm pk,obs}$ XRFs as well as a potentially large missing population of short-duration GRBs that will be probed by EXIST.
\end{abstract}


\maketitle

\section{Introduction}
\label{sec:intro}

The {\it Swift}~satellite \citep{gehrels04} has transformed the study of
Gamma-ray Bursts (GRBs) and their afterglows.  Our knowledge of the early X-ray afterglows
has increased tremendously due to the dramatic success of the X-ray Telescope \citep[XRT;][]{burrows05}. However,
our understanding of the prompt emission properties has lagged.  This is due in part
to the narrow energy bandpass of the Burst Alert Telescope \citep[BAT;][]{bart05}, which precludes direct
measurement of the broad GRB spectra and tends to weaken any inferences about the $\nu F_{\nu}$ spectral
peak energy $E_{\rm pk,obs}$ and the bolometric GRB fluence.  

In the first installment of our ``Complete BAT Catalog
of Swift GRBs and Spectra'' \citep[][hereafter Paper I]{butleretal07}, we treat these limitations of the
BAT in a statistically rigorous fashion and study tantalizing
pre-Swift correlations between the host-frame
characteristics of GRBs \citep[e.g.,][]{lpm00,fr00,nmb00,shaf03,amati02,lamb04,ggl04,firmani06}.
A number of these potential log-log relations
appear dramatically different in the Swift-era sample, with a broader scatter, and a shift in normalization toward the detector threshold.
From this, we concluded that the origin of these correlations was 
tied more closely to the detection process than to the intrinsic physics of GRBs.

We present here fits to the lightcurves and spectra of additional Swift GRBs detected between April 17, 2007 and August 13, 2009,
nearly doubling the overall sample.  As summarized in Paper I and below, our analysis is extremely uniform (nearly fully-automated), with well-defined
survey flux limits,  and our fits allow for detailed propagation of errors.  These features are critical to the estimate of GRB rates,
the focus of the current work.  Our primary goal is to uncover, with realistic estimates of uncertainty, the intrinsic GRB production rate
as a function of redshift $\dot \rho(z)$.  To measure this quantity, it is necessary to model the GRB luminosity function $\phi(L)$
allowing for possible intrinsic and detection-based correlations.  

\begin{figure} 
\hspace{-0.2in}
\includegraphics[width=3.7in]{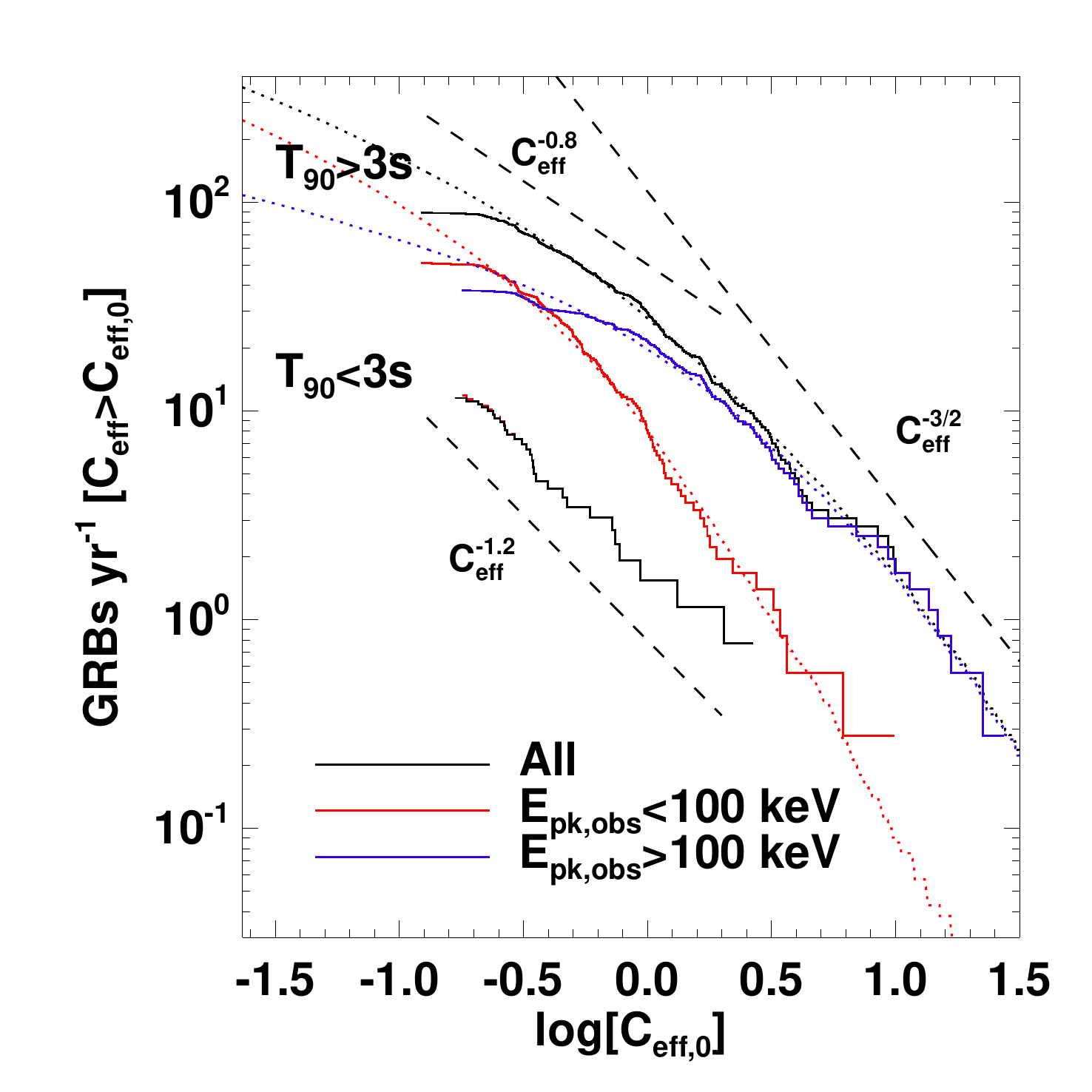} 
\caption{\small 
Strong spectral and temporal dependence in the number of GRBs with effective count rate $C_{\rm eff}$ (Section \ref{sec:overview}) 
above a given value (i.e., the Swift ``logN--logS'' curve; see Section \ref{sec:prior}).  We plot here long duration ($T_{90}>3$ s) 
GRBs with hardness above and below the median Swift $E_{\rm pk,obs}= 100$ keV and also
short duration ($T_{90}<3$ s) GRBs.  The rate of long-duration hard GRBs is turning over at low 
flux levels, while the rate of long-duration soft GRBs rises more strongly.  This is a gradual effect in $E_{\rm pk,obs}$.
Although the logN--logS slope for long-duration GRBs does not appear to be duration dependent, the Swift short-duration GRB 
population is strongly rising in number to low flux levels, showing no significant sign of a turn-over.
The curves expected without
a cutoff --- derived in Section \ref{sec:fitting} --- due to the detector are plotted as dotted lines.
The dashed red curve (barely visible) at the left of the short-duration curve accounts for the detector threshold
following the non-parametric prescription of \citet{pet96}.
} 
\label{fig:logN-logS} 
\end{figure}

We derive a model that describes both Swift and pre-Swift rates well as a function of hardness, duration, flux, and redshift (Section \ref{sec:redux}).
As we discuss in Section \ref{sec:fitting} below, we find a highly significant, intrinsic 
correlation between $E_{\rm iso}$ and $E_{\rm pk}$;
however, the observed correlation has large scatter and is strongly instrument-dependent.
Figure \ref{fig:logN-logS} shows how the overall number of events observed by Swift --- and not just the correlation 
between observed quantities --- requires us to build temporal and spectral dependences into modelling the luminosity function.  

In the next section, we summarize the utility and historical background behind making a
plot like Figure \ref{fig:logN-logS}.
We then discuss in Section \ref{sec:fitting} the models that successfully recreate the curves in Figure \ref{fig:logN-logS} and also the implications for the GRB luminosity function (Section \ref{sec:energetics}) and
comoving rate density (Section \ref{sec:rho}).
In Section \ref{sec:exist} we make self-consistent predictions for the expected observed redshift distribution for all Swift GRBs --- including the
60\% fraction of Swift GRBs for which a spectroscopic $z$ has not been measured --- and also for the planned and more-sensitive EXIST experiment.
We expect EXIST to thrive with respect to the detection of both short and long duration GRBs at high redshift.

\subsection{Prior Rate and Luminosity Function Estimates}
\label{sec:prior}

There is a rich literature describing optimal ways of counting GRBs to determine their distance and intrinsic flux.
In the pre-afterglow era, counting focused on the observed flux distributions.
The number of events $N$ with observed flux greater than $S$ --- the so called ``logN--logS'' curve --- showed
early evidence (over many decades in $S$) for slope $S^{-3/2}$ expected for a homogeneous, isotropic, and static source population in
a Euclidean universe \citep[HISE; e.g.,][]{hurley91,hs90}.   A powerful statistic for examining the source counts
is $V/V_{\rm max} = (C/C_{\rm min})^{-3/2}$ \citep{schmidt68}, a measure of the volume probed by a source detected with $C$ counts relative to a possible
minimum number of observable counts $C_{\rm min}$.  The expectation is that $\langle V/V_{\rm max} \rangle = 0.5$ for HISE.  The
BATSE experiment provided the first strong evidence from a single experiment \citep{meegan92} --- a deficit of low $S$ GRBS and $\langle V/V_{\rm max} \rangle <0.5$ --- 
for a departure from homogeneity, while the spatial counts showed clearly an isotropic population.
To study whether these modest departures imply that GRBs are very local (a Galactic halo population) or cosmological required examination
beyond the first moment $\langle V/V_{\rm max} \rangle$ in the $V$ (or $C$) distribution  \citep[e.g.,][]{band92,ht93,pet93}.

The first GRB redshifts \citep[e.g.,][]{metzger97} defined a cosmological origin and a vast energy release.
Connecting the small number of GRBs with $z$ to the large population of GRBs without $z$ required, in general, careful modelling
of and strong assumptions for the intrinsic luminosity and number density distributions in order to reproduce the observed flux data 
\citep[e.g.,][]{piran92, piran99, cp95, fb95, lw95, lw98, hh97, schmidt99, schmidt01, sb01, guetta05}.
Exceptions to the parametric approach were studies utilizing luminosity criteria (i.e., possible correlations of observables with luminosity)
to derive ``pseudo-redshifts'' for the full GRB sample \citep[e.g.,][]{norris00,fr00,schaf01,lloyd02,mur03,yon04,firm04,kocevski06,schmidt09}.  These studies generally found a rising GRB rate to $z \lessim 2$, similar to
the cosmic star formation rate \citep[e.g.,][]{madau96}, but potentially continuing to remain flat or even rising
to $z\sim 12$.  

Notably, some of these works \citep[e.g.,][]{lloyd02,yon04}, using hazard statistics \citep{lb71,ep92,pet93,mp99},
also found evidence for potential strong luminosity evolution, parameterized as $L\propto (1+z)^a$, with $a$ in the range 1.5--2.5.  
The luminosity function itself appears to generally be characterized well
as broken powerlaw, with a break at $L\sim 10^{51-52}$ erg s$^{-1}$ and a flat or slowly rising slope to low-energies, strongly dependent upon the instrumental detection model.

The connection between GRBs and the deaths of massive stars (now firmly established, e.g., \citet{stanek03,hjorth03}; see \citet{woosbloom06} for a review) sped progress by motivating an assumption that the GRB
rate follows star formation \citep[e.g.,][]{wijers98,lr00,pm01,cs02,bloom03,gor04,nat05}.
Very recently, thanks to Swift and the impressive efforts of ground-based observers, a growing sample of GRBs with spectroscopic redshifts
has allowed for direct tabulation of GRB intrinsic luminosities \citep[e.g.,][]{kocevskibutler08} and 
redshifts \citep[e.g.,][]{jak05,jak06}.

The large number of redshifts has also enabled a detailed comparison of the intrinsic GRB rate to the cosmic star formation rate \citep[SFR, e.g.,][]{daigne06,salv07,kistler08,kistler09,salv09a,salv09b}.  Perhaps the most intriguing, shared feature of these studies is a strong indication of evolution in the GRB population.  Above $z\approx 2$
and possibly extending to $z\approx 8$, the Swit GRB rate is increasing far faster than star formation \citep[e.g.,][]{kistler08,kistler09}, and it is not clear to what extent this is due to
GRBs in the early universe being bright \citep[i.e., luminosity evolution, preferred by][]{salv07,salv09a,salv09b,pet09} or to an increase in the overall number of GRBs 
at intermediate and high-$z$ relative to the SFR.

As we discuss below in Section \ref{sec:rho}, rigorous treatment of the largest available Swift dataset (Section \ref{sec:redux}) allows for a firm conclusion in favor of rate evolution
and not luminosity evolution, and we suggest plausible explanations.  To draw this conclusion and to study GRB rates as a function of intrinsic hardness, flux, and duration (Sections
\ref{sec:fitting} \& \ref{sec:energetics}) as well as $z$, we require a detailed model for the Swift satellite detection limit (Section \ref{sec:limit}).

\section{Data Reduction and Fits}
\label{sec:redux}

Our automated pipeline at the University of California, Berkeley is used to download the \swift~data in near real
time from the {\it Swift}~Archive\footnote{ftp://legacy.gsfc.nasa.gov/swift/data}
and quicklook site.  
We use the calibration files from the 2008-12-17 BAT database release.
We establish the energy scale and mask weighting for the BAT event mode data 
by running the {\tt bateconvert} and {\tt batmaskwtevt} tasks
from the HEASoft 6.6.1 software release\footnote{http://swift.gsfc.nasa.gov/docs/software/lheasoft/download.html}.  
Spectra and light curves are extracted with the {\tt batbinevt} task, and response
matrices are produced by running {\tt batdrmgen}.  We apply the systematic
error corrections to the low-energy BAT spectral data as suggested by the BAT
Digest website\footnote{http://swift.gsfc.nasa.gov/docs/swift/analysis/bat\_digest.html} \citep[see, also][]{taka08}, and fit the data in
the 15--150 keV band using 
ISIS\footnote{http://space.mit.edu/CXC/ISIS}.  The spectral normalizations are corrected for satellite slews as recommended in BAT Digest.
All errors regions reported correspond to the 90\% confidence interval.  In determining source
frame flux values, we assume a cosmology with $h=0.71$, $\Omega_m=0.3$, and $\Omega_{\Lambda}=0.7$.

The burst duration intervals are determined automatically as described in Paper I and are presented in
Table 3 for the current, supplemental sample.   Spectral fitting is perform also
as described in Paper I, and we present the results in Table 4.  Electronic version of the tables --- updated in near real time ---
including additional fit statistics (e.g., fluxes and fluences
in various bandpasses) and downloadable reduced data, can be found at the project webpage\footnote{http://astro.berkeley.edu/$\sim$nat/swift}.

\subsection{Sample Selection \& Survey Flux Limit}
\label{sec:limit}

Although Tables 3 \& 4 contain data for the full Swift sample, we restrict the rate analysis below to long-duration GRBs \citep[$T_{90}>3$ s, e.g.,][]{kouv93} with signal-to-noise ratios  $S/N>10$.  The separation of short and long duration GRBs is motivated by the potential that the duration classes map to separate source populations
\citep[e.g.,][and references therein]{zhang08,nys09,lev09}, also suggested by the sharp logN--logS slope variation (Figure \ref{fig:logN-logS}) for short GRBs relative to long GRBs.
We exclude 3 GRBs (051109B, 060218, and 060614) at $z<0.2$ due to data quality issues (uncertain
redshift, missing data, possible short duration, respectively; see Paper I).  We note that the $\approx 6$\% of BAT GRBs detected in ground analyses are not included in this catalog.

Central to our analysis is a burst-by-burst estimate of the minimum detectable count rate.  As discussed in Paper I, an a-posterior estimate of the optimal imaging $S/N$ for every GRB detected by BAT can be obtained by generating the demasked light curve using the XRT position for the GRB.  The temporal region which maximizes the $S/N$ can be found, and this maximal $S/N$ (see Table 1) bounds that achievable by the BAT trigger software.  The maximal $S/N$ can be used to infer the minimum detectable counts $C_{\rm min}$ relative to the observed number of counts $C$:  $C_{\rm min} = C (10/[S/N])$ ($10\sigma$ detection limit).  This is a valid approximation only in the background-dominated noise regime, and there is a modest $\approx 10$\% correction in the case of a handful of the very brightest Swift GRBs.  

This $C_{\rm min}$ estimate effectively treats the BAT trigger software as perfect, always able to find the optimal $S/N$ region.  This approximation should break down in the limit of low $S/N$, which is why we have chosen $10\sigma$ (as opposed to, e.g., $5\sigma$).  The observed $C/C_{\rm min}$ distribution turns over below $10\sigma$, justifying this choice.

As discussed in detail in Paper I, the photon fluence over root duration correlates strongly with our $S/N$ measure, indicating that the BAT instrument sensitivity is best characterized in such units \citep[see, also,][]{taka07}.  We can define an effective count rate $C_{\rm eff} = C \sqrt{ f_p/T_{r45} }$, where $f_p$ is the partial coding fraction resulting from the position of the GRB in the BAT field-of-view, and $T_{r45}$ is the high-signal time duration of \citet{reichart01}.  This duration measure should provide a good link between the time-integrated flux and the flux --- while the GRB is bright --- relevant for triggering.  The quantity $C_{\rm eff}/[S/N]$ clusters tightly with a scatter $\lessim 0.1$ dex, indicating an effective BAT threshold count rate of $C_{\rm eff,min} = 0.24\pm 0.05$ cts/s$^{0.5}$/fully-coded-detector ($10\sigma$).  We interpret the uncertainty in this number as the uncertainty in our ability to measure the true threshold count rate given our estimate of that rate.  

It is a very poor approximation, alternatively, to model Swift as a peak photon flux detector like BATSE.  Swift GRBs with $S/N<10$ have peak photon fluxes in the 15--150 keV band ranging from
0.1 to 1 ph cm$^{-2}$ s$^{-1}$.  It would be necessary to discard 40\% of the full sample (that with peak fluxes below 1 ph cm$^{-2}$ s$^{-1}$) to obtain a clean, flux-limited sample.  Worse yet, a $10\sigma$ threshold based on 15--150 keV photon fluence over $T_{90}$ duration would need to reject $\gtrsim 60$\% of the Swift sample to avoid ($p<0.6$ cm$^{-2}$ s$^{-1}$) GRBs whose numbers are strongly affected by the detection limit.  Our approach allows for fitting of 88\% of the full sample at 10$\sigma$.

\subsection{GRB World Model Overview}
\label{sec:overview}


We wish to derive a model capable of reproducing the observed Swift GRB rate as a function of redshift, flux, hardness, and duration.  As discussed above,
these quantities are known to --- or have been argued to --- exhibit strong correlations. Therefore,  all must be considered in deriving reliable rates.  Additional quantities (e.g., to describe GRB beaming)
may be important but are not readily accessible in order to be grafted onto the present catalog.
Our formalism
rigorously accounts for measurement errors and for correlations present in the data.

We characterize the GRB rate as a product of terms involving the redshift $z$ of the bursts, the isotropic equivalent energy release (1--$10^4$ keV)
$E_{\rm iso}$, the duration $T_{r45}$, and also the hardness $E_{\rm pk,obs}$ of the bursts.  
To describe correlations among these quantities, it is sufficiently general (see Appendix) to define an effective bolometric luminosity $L$:
\begin{equation}
L = {E_{\rm iso} \over (1+z)^{\alpha_z}} \left({10^{2.5} {\rm keV} \over E_{\rm pk}} \right)^{\alpha_E} \left( {10^{0.6} {\rm s}\over T_{r45,z}} \right)^{\alpha_T}.
\label{eq:leff}
\end{equation}
We assume a smoothly broken powerlaw for the luminosity function:
\begin{equation}
\phi_L = {dN \over d\log{L}} = \begin{cases}
  (L/L_{\rm cut})^{a_L}  & L<L_{\rm cut} \\
  (L/L_{\rm cut})^{b_L}  & L \geq L_{\rm cut}. \\
\end{cases}
\label{eq:lum_func}
\end{equation}
We make this function smooth by convolving it with a Gaussian in $\log{(L)}$ of width $\sigma_L$, resulting in $\phi(L) \rightarrow \tilde \phi(L|\sigma_L)$.
In the case of no luminosity correlations ($\alpha$'s all zero), $\tilde \phi$ is the distribution for $E_{\rm iso}$.

The true, detector-independent differential rate (per $z$, per
$\log{E_{\rm iso}}$, $\log{E_{\rm pk}}$, $\log{T_{r45z}}$)
can then be written:
\begin{equation}
r_{\rm true} = \tilde \phi(L) P_E(E_{\rm pk}) P_T(T_{r45,z}) {r_0 \dot \rho(z) dV/dz \over (1+z)},
\label{eq:rtrue}
\end{equation}
where $P_E(E_{\rm pk})$ is the intrinsic $E_{\rm pk}$ distribution, $P_T(T_{r45,z})$ is the intrinsic distribution in $T_{r45,z}$, and $\dot \rho(z)$
is the comoving GRB rate density.  The universal volume is $V$, and the factor $(1+z)$ accounts for cosmic time dilation.  The normalization $r_0 \propto dt~d\Omega~/\langle f_{\rm beam} \rangle$ includes the survey duration, solid angle $d\Omega\approx 1.4$ sr, and the effects of GRB beaming.
Functional forms for $P_E(E_{\rm pk})$, $P_T(T_{r45,z})$,
and $\dot \rho(z)$ are described below.

For a given GRB, the expected number of BAT counts $C$ (Section \ref{sec:redux}) in the trigger time window depends on $L$, $E_{\rm pk}$, $T_{\rm r45,z}$, and $z$.  The GRB will be detected
when $C>C_{\rm min}$, where $C_{\rm min}$ depends approximately on $T_{r45}$ alone, yielding the observed GRB rate:
\begin{equation}
r_{\rm obs}= \Theta(C-C_{\rm min}) ~r_{\rm true}.
\label{eq:robs}
\end{equation}
We fit this multivariate model (Section \ref{sec:fitting}) by evaluating and maximizing $r_{\rm obs}$ at each observed data point
(see Appendix for more details).

\subsection{The GRB Rate Density}

Based on the shape of the cosmic star formation rate \citep[SFR;][]{hopkinsbeakom}, we assume a broken powerlaw for the comoving GRB rate density:
\begin{equation}
 \dot \rho(z) = {dN \over dz} \propto \begin{cases}
 (1+z)^{g_0}  & z<0.97 \\
 (1+z)^{g_1} & 0.97<z<z_1 \\
 (1+z)^{g_2} & z>z_1, \\
            \end{cases}
\label{eq:rho}
\end{equation}
where the relative normalizations (not written above) are set so that $\dot \rho(z)$ is continuous at $z_0=0.97$ and $z_1$.  The SFR has roughly ($g_0$, $g_1$, $g_2$) $=$ (3.4, $-$0.3, $-$8) for $z_1\approx 4.5$ (e.g., Figure \ref{fig:rate_dens}).  For this model to yield acceptable fits to
the Swift GRB data (Section \ref{sec:fitting}), the parameters $g_0$, $g_1$, $g_2$, and $z_1$ must be allowed to vary.  In fitting this
model, we marginalize over the free parameters and derive a best-fit shape that is generally smoother than the input
form.  Future work can compare star formation models directly to these best-fit curves and error regions without needing to re-fit the Swift data.

\subsection{The Intrinsic $E_{\rm pk}$ Distribution}

We have tested a variety of functional forms for the intrinsic distribution in $\log[E_{\rm pk}]$.  A normal distribution can fit the observed univariate
distribution reasonably well, but it fails to account well for the multivariate distribution in $E_{\rm pk}$ and $E_{\rm iso}$ (Section \ref{sec:dists}).  This is because
a narrow log-Gaussian cannot generate a large enough dynamic range in $E_{\rm pk}$ to allow for the observed correlation (over many decades in $E_{\rm pk}$) with $E_{\rm iso}$.
Moreover, the observed $E_{\rm pk}$ distribution is found also to be consistent with intrinsic models formed from the sum of log-Gaussian extending to low
$E_{\rm pk}$.  That is, low $E_{\rm pk}$ GRBs have a tendency to go undetected, and the data do not strongly constrain the presence of such populations.  

To fit
a more general form for the intrinsic $\log[E_{\rm pk}]$ distribution, we consider the following:
\begin{equation}
P_E = {dN \over d\log{E_{\rm pk}}} = \begin{cases}
  (E_{\rm pk}/E_{\rm pk,0})^{\beta_E} & E_{\rm pk}<E_{\rm pk,0} \\
  \exp[-{1 \over 2} ({\log{(E_{\rm pk}/E_{\rm pk,0})} \over \sigma_{E_{\rm pk,0}}})^2] & E_{\rm pk}>E_{\rm pk,0}. \\
  \end{cases}
\end{equation}
This functional form allows for a potential population of low $E_{\rm pk}$ events not readily detected by Swift.  We find that the low-$E_{\rm pk}$ powerlaw
index $\beta_E$ tends to be negative, indicating a large number of missing low $E_{\rm pk}$ events.
The proposed extent of such a population is not new to
this work \citep[e.g.,][]{stroh98,lamb05,pel08}.

\subsection{The Intrinsic $T_{r45,z}$ Distribution}

The intrinsic duration distribution $P_T(T_{r45,z})$ appears to be well-modelled as a log-Gaussian with variable mean and width:  
\begin{equation}
P_T = {dN \over d\log{T_{\rm r45,z}}} =
  \exp[-{1 \over 2} ({\log{(T_{\rm r45,z}/T_{\rm r45,z0})} \over \sigma_{T_{\rm r45z,0}}})^2].
\end{equation}
There is evidence for a modest
fraction of missing long-duration events (Figure \ref{fig:fits}); however additional components to $P_T(T_{r45,z})$ do not appear to be necessary to model this.
We note that we use the common transformation between observed duration $T_{r45}$ and intrinsic duration $T_{r45,z} = T_{r45} / (1+z)^{0.6}$ \citep[e.g.,][]{firmani06},
which accounts for the expected broadening of pulse widths due to spectral evolution observed at softer bandpasses for GRBs at higher $z$ \citep{fen05}.

\subsection{Model Fitting}
\label{sec:fitting}

We fit the rate density model (Equation \ref{eq:robs}) to the data for 120 Swift GRBs with measured redshift and 205 additional GRBs without measured redshift.
We assume uniform priors for all model parameters summarized in Table 1.
The fitting is accomplished by maximizing Equation \ref{eq:logl} using Markov Chain Monte Carlo (MCMC) in python with PyMC\footnote{http://code.google.com/p/pymc}.

We have performed the fitting for two sample divisions: (1) GRBs with redshift only, and (2) GRBs with and without redshift.
The best-fit parameters after including GRBs without redshift are closely consistent ($1\sigma$ level) with those
found from considering only the GRBs with redshift.  This indicates that Swift GRBs without redshift do not have a strongly different redshift distribution
as compared to the GRBs with redshifts.  In any case, the analysis below does not depend strongly on the inclusion/exclusion of Swift GRBs without measured redshift
(see, Section \ref{sec:redshift}), apart from the estimate of total expected rates from future experiments (Section \ref{sec:exist}).
We will focus on the parameters derived for the full sample, because the redshift-only sample is, in principle, more prone to biases related to obtaining spectroscopic
redshifts (Section \ref{sec:redshift}).

\begin{figure*}
\center{\includegraphics[width=3.2in]{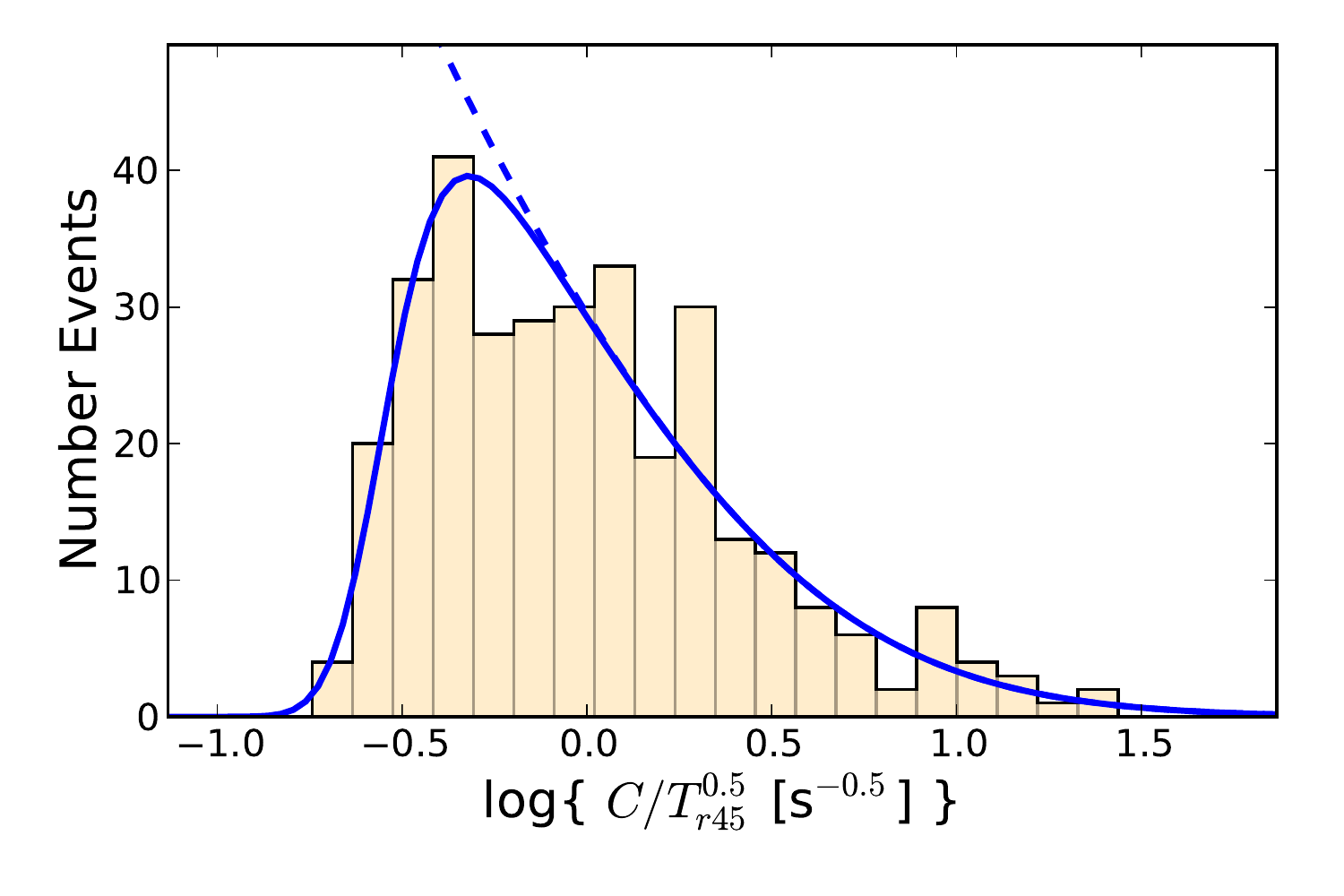}
\includegraphics[width=3.2in]{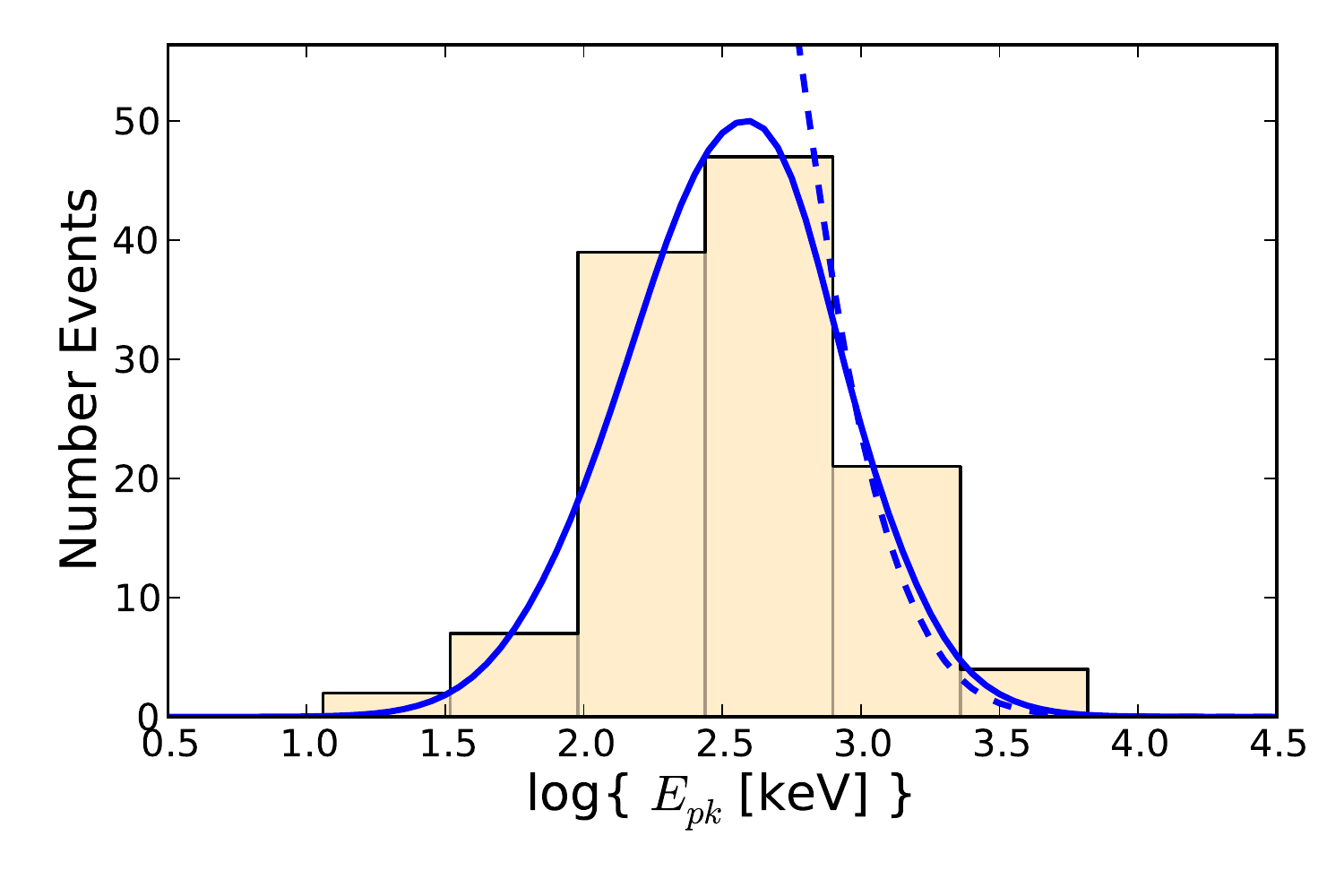}}
\vspace{-0.25in}
\center{\includegraphics[width=3.2in]{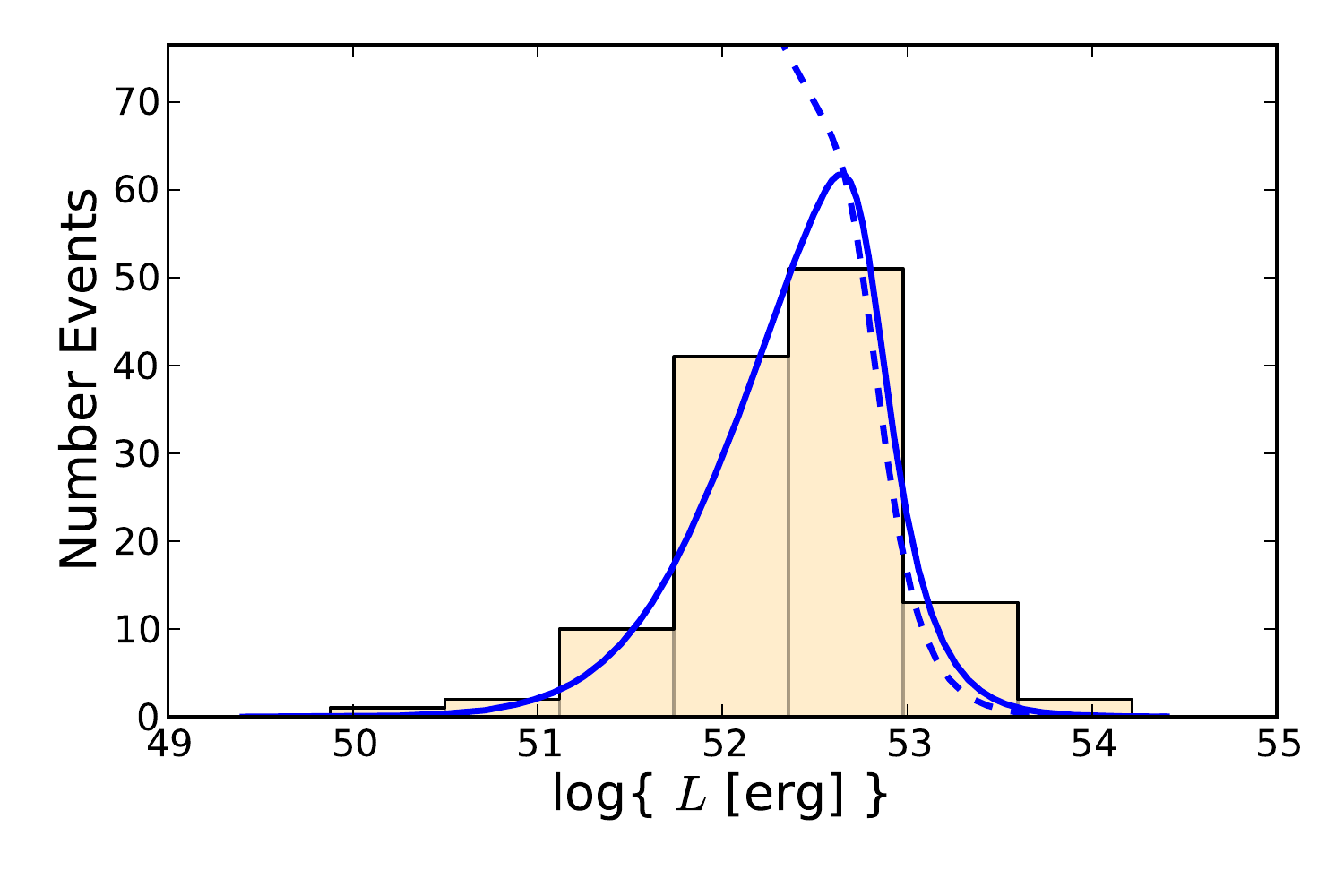}
\includegraphics[width=3.2in]{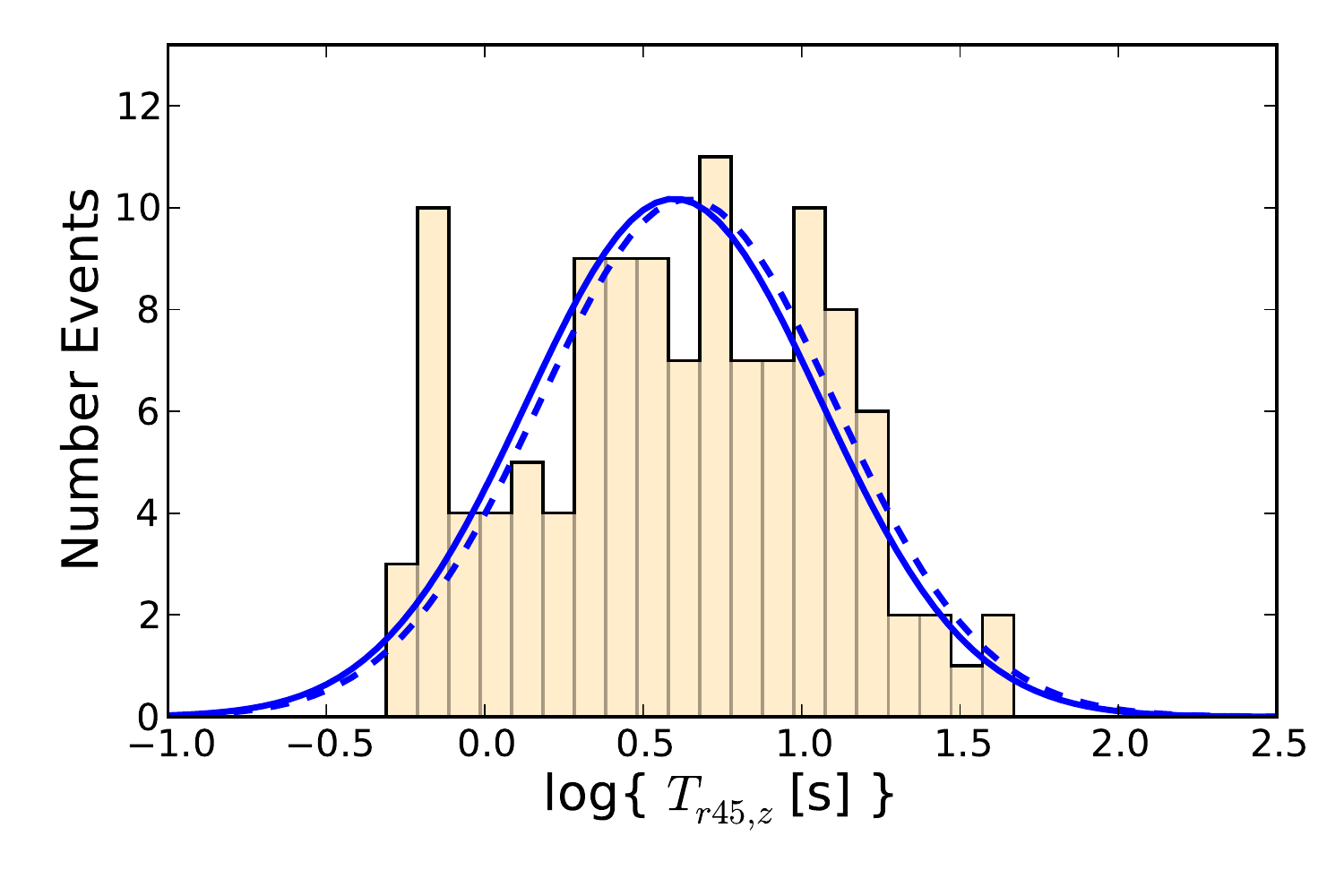}}
\vspace{-0.25in}
\center{\includegraphics[width=3.2in]{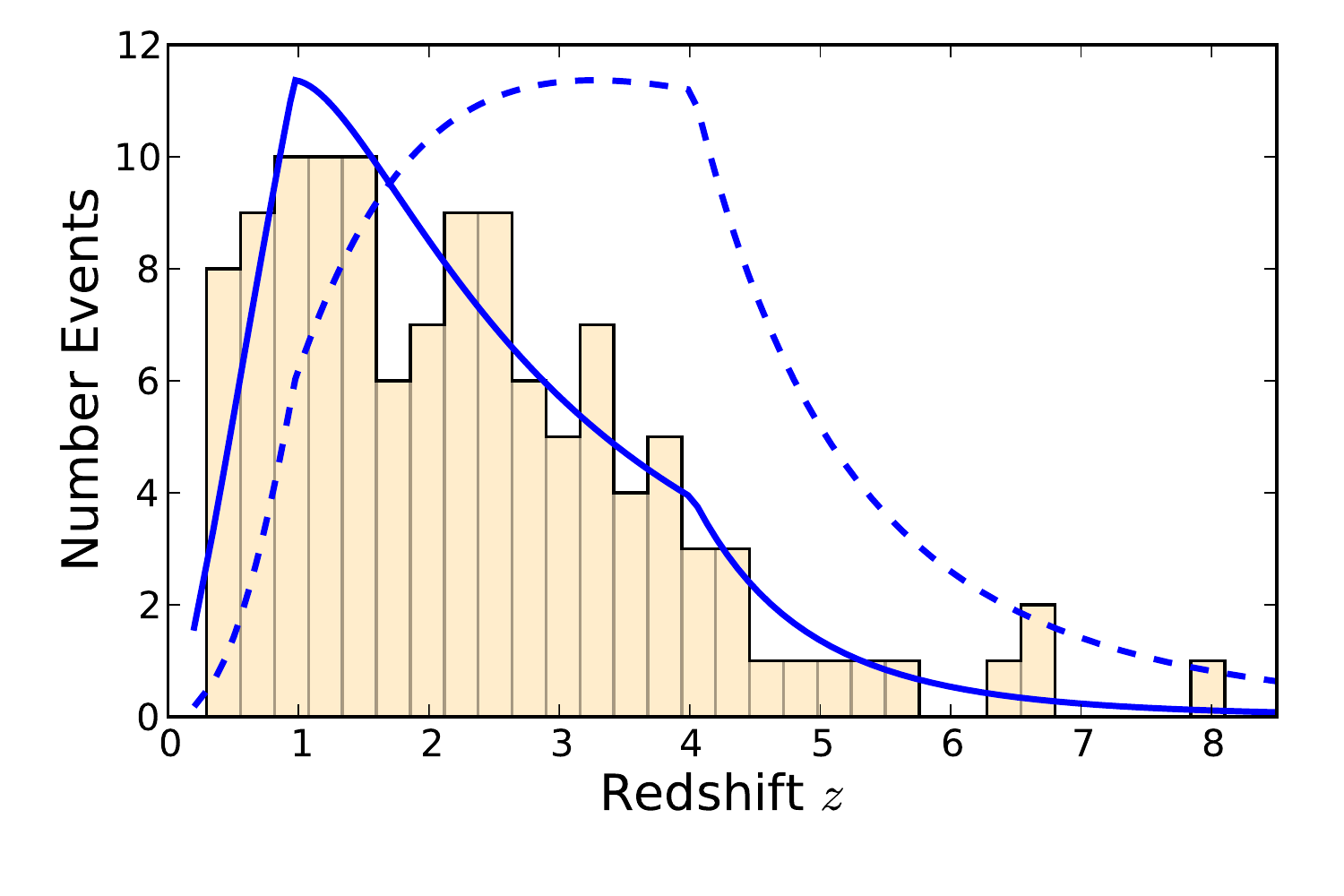}}
\vspace{-0.2in}
\caption{\small 
The best-fit GRB world model from Table 1 (left) superposed (solid lines) on the Swift data: 
counts $C$, redshift $z$, duration $T_{r45,z}$, hardness $E_{\rm pk}$, and effective luminosity $L$ (Section \ref{sec:overview}) 
The input models prior to convolution through the detector are plotted as dashed lines.
The measured values for $E_{\rm pk}$ and $L$ have large errors, with typical sizes that correspond to the bin widths used in the $E_{\rm pk}$ and $L$ sub-panels.
For all but the first sub-panel, the data and models are for GRBs with measured redshift only.  
Each curve provides an excellent fit to the data (Section \ref{sec:fitting}).  
Note the large number
of ``missing'' GRBs with low $C$, $E_{\rm pk}$, and $L$.} 
\label{fig:fits} 
\end{figure*}

Figures \ref{fig:fits} show the best fit model from Table 1 (including GRBs without redshift) overplotted on the univariate data distributions.  The quality of
the fits can be judged visually or from a Kolmogorov-Smirnov (KS) test \citep[e.g.,][]{press92}.  In particular, the KS-test null hypothesis probability that the observed redshift distribution
is different than the predicted distribution is $P_{\rm KS}=0.94$, indicating little evidence that the model and data distributions significantly differ.  The KS-test, likewise, for the
distribution in detected counts $C$ yields $P_{\rm KS}=0.29$.   For the $T_{r45,z}$, $L$, and $E_{\rm pk,z}$ distributions we find $P_{\rm KS}=0.77$, $P_{\rm KS}=0.77$, and $P_{\rm KS}=0.84$, respectively.  
We have also performed two-dimensional KS tests \citep[e.g.,][]{press92} on the bivariate distributions of
$E_{\rm iso}$ with $E_{\rm pk}$ ($P_{\rm KS}=0.17$; see Figure \ref{fig:epeiso_pred}), or $T_{r45,z}$ ($P_{\rm KS}=0.12$), or $z$ ($P_{\rm KS}=0.14$), indicating no strong evidence for a poor fit.  Considering all of the
KS-tests above, the model appears to reproduce
the data quite well.

We note that an intrinsic correlation between $E_{\rm iso}$ and $E_{\rm pk}$ is present ($\alpha_E=1.8\pm0.3$) and is highly significant
($\alpha_E>0$ at $9\sigma$, $-2\Delta \log(\mathcal{L})=82.6\sim \Delta \chi^2$ for one additional degree of freedom).  Some of
this correlation may be with the
variables $T_{r45,z}$ and $z$; however, there is little evidence that luminosity evolution ($\alpha_z=0.0\pm 0.5$) is present, and the
evidence for correlation between $E_{\rm iso}$ and $T_{r45,z}$ ($\alpha_T = 0.40\pm 0.2$, full sample; $\alpha_T=0.3\pm 0.2$ $z$ only) is 
comparatively weak.  

We note that the magnitude (i.e., $\alpha_T$) of the intrinsic correlation with $T_{r45,z}$ covaries strongly with the center
of the intrinsic $T_{rt45,z}$ distribution $\log{(T_{rt45,z0})}$ (Pearson correlation coefficient $r=-0.74$), which is already displaced toward long $T_{r45,z}$
from the observed distribution (Figure \ref{fig:fits}).  To the extent that we have potentially over-estimated the Swift sensitivity at long durations by assuming that the trigger
sensitivity scales indefinitely as $T_{r45}^{-0.5}$, the evidence for an intrinsic correlation between $E_{\rm iso}$ and $T_{r45,z}$ would
weaken.

\begin{figure} 
\includegraphics[width=4.5in]{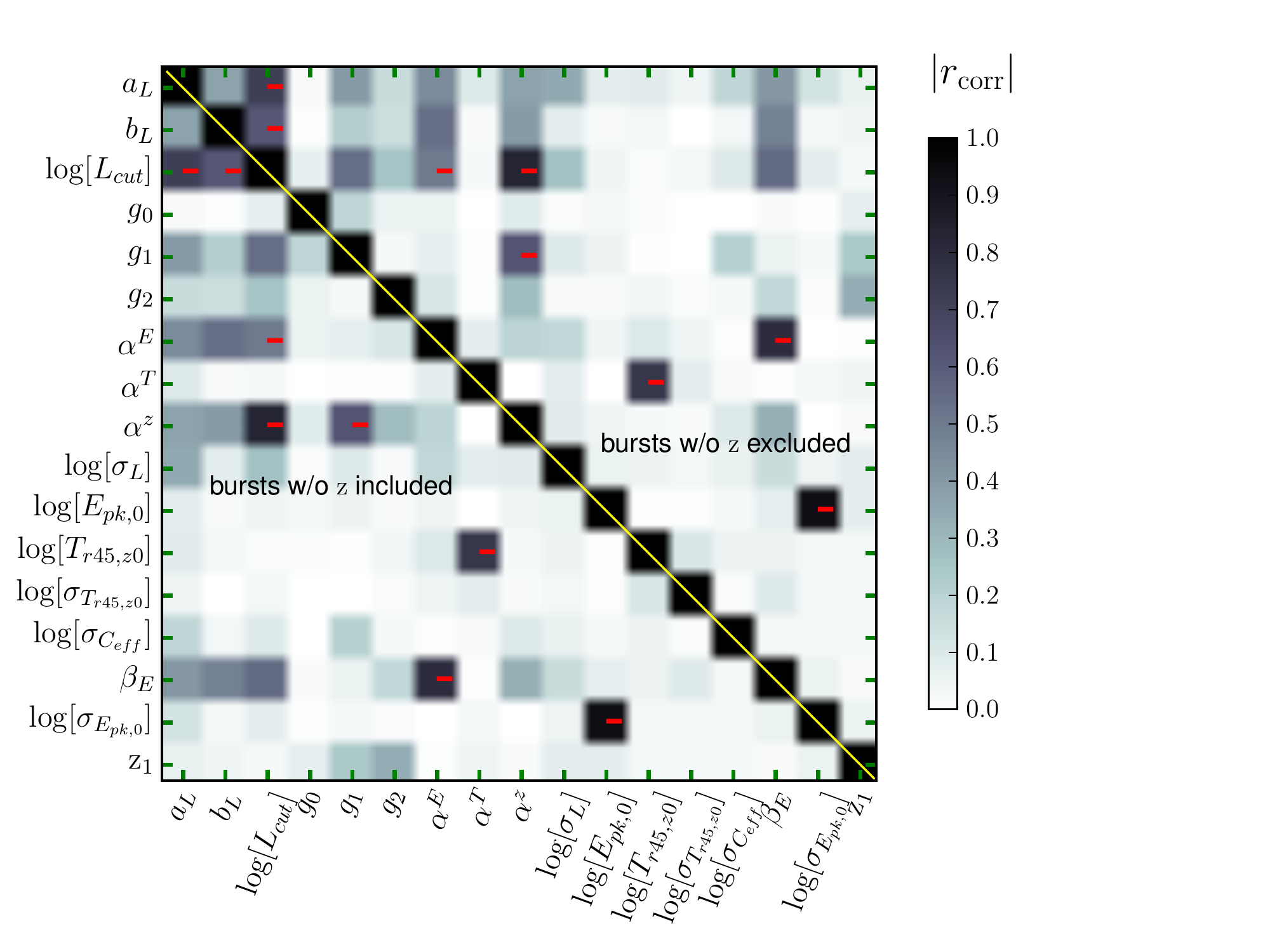}
\caption{\small 
As demonstrated by the sample correlation matrix,  the parameters are generally not 
strongly correlated (see, Section \ref{sec:fitting}), thanks largely to the existence of measured redshifts. 
Here, we plot the matrix for all GRBs in the sample in the lower diagonal,  
and we plot the matrix for just GRBs with redshifts in the upper diagonal. Negative entries are marked with 
a red ``-''.} 
\label{fig:corr} 
\end{figure}

Figure \ref{fig:corr} displays the sample correlation matrix.  In addition to the covariance just mentioned for $\alpha_T$ and $\log{(T_{rt45,z0})}$, there are
a number of additional parameters which strongly co-vary (Pearson correlation coefficient $r>0.5$).  Covariance is minimized when ignoring the GRBs without
measured redshifts:  the value of $\log{(L_{\rm cut})}$ and the strength of possible luminosity evolution $\alpha_z$ exhibit
$r=-0.74$, and the center of the intrinsic distributions in $E_{\rm pk}$ correlates with the width of the distribution ($\log{(\sigma_{E_{\rm pk,0}})}$)
with $r=-0.90$.  

It is important to note that the parameters describing the rate density do not co-vary strongly with the other parameters.
This is not true when the sample without redshift is included in the analysis.  In that case, the slope of $\dot \rho$ between
$z=0.97$ and $z\approx 4$ covaries with $\log{(L_{\rm cut})}$ ($r=0.54$) and with the evolution index $\alpha_z$ ($r=-0.63$).
Finally, the $E_{\rm iso}-E_{\rm pk}$ correlation index $\alpha_E$ anti-correlates strongly with the parameter $\beta_E$ describing the
rate of missing low-$E_{\rm pk}$ GRBs, particularly in the case where the GRBs without redshifts are included ($r=-0.80$).

\begin{table}[htbp]
\caption{Best-fit GRB World Models}
\begin{tabular}{l|c|c}\hline\hline
parameter & value (all GRBs) & value (w/ $z$ only) \\
& & \\
Luminosity Func.  & & \\\hline

$a_L$ & $-0.22$ ($-0.18$,$+0.31$) & $-0.27\pm0.19$ \\
$b_L$ & $-2.89$ ($-2.05$,$+1.06$) & $-3.46\pm1.53$ \\
$\log{[L_{\rm cut}]}$ & 52.74 $\pm$ 0.43 & $52.95\pm0.31$ \\
$\log{[\sigma_L]}$ & $-1.57 \pm 1.16$ & $-1.78 \pm 1.12$ \\

& & \\
Rate Density & & \\\hline
$g_0$ & $3.14 \pm 0.71$ & $3.35 \pm 0.74$ \\
$g_1$ & $1.36 \pm 0.58$ & $1.32 \pm 0.58$ \\
$g_2$ & $-2.92$ ($-2.36$,$+1.58$) & $-2.51$ ($-2.25$,$+1.60$) \\
$z_1$ &  $4.00 \pm 0.38$ & $3.98 \pm 0.38$ \\

& & \\
$E_{\rm iso}$ Correlations & & \\\hline
$\alpha_E$ $(E_{\rm pk}$) & $1.77 \pm 0.28$ & 1.59 ($-0.24$,$+0.32$) \\
$\alpha_T$ $(T_{r45,z})$ & $0.40 \pm 0.15$ & $0.27 \pm 0.22$ \\
$\alpha_z$ $(z)$ & $0.05 \pm 0.49$ & $-0.12 \pm 0.45$ \\

& & \\
$E_{\rm pk}$ dist. & & \\\hline
$\log{[E_{\rm pk,0}]}$ &  2.60 ($-0.42$,$+0.27$) & 2.60 ($-0.51$,$+0.33$) \\
$\beta_E$ & $-0.51 \pm 0.47$ & $-0.29$ ($-0.45$,$+0.69$) \\
$\log{[\sigma_{E_{\rm pk,0}}]}$ & $-0.65 \pm 0.26$ & $-0.57$ ($-0.34$,$+0.25$) \\

& & \\
$T_{r45,z}$ dist. & & \\\hline
$\log{[T_{r45,z0}]}$ & $0.63 \pm 0.07$ & $0.68 \pm 0.11$ \\
$\log{[\sigma_{T_{rt45,z0}}]}$ &  $-0.33 \pm 0.03$ & $-0.32 \pm 0.05$ \\

& & \\
Threshold & & \\\hline
$\log{[C_{\rm eff,min}]}$ & $-0.59 \pm 0.01$ & $-0.59 \pm 0.02$ \\
& & \\\hline
\end{tabular}
\label{tab:pars}
\end{table}

\subsection{Pre-Swift Distributions}
\label{sec:dists}

To gauge the validity of the model fit above to the Swift sample, we can compare the model predictions to data obtained from pre-Swift experiments.
We do this here for the hardness and flux distributions. Studies of the duration distributions \citep[e.g.,][]{lev09}
also appear to demonstrate consistency.

\begin{figure} 
\hspace{-0.2in}
\includegraphics[width=3.7in]{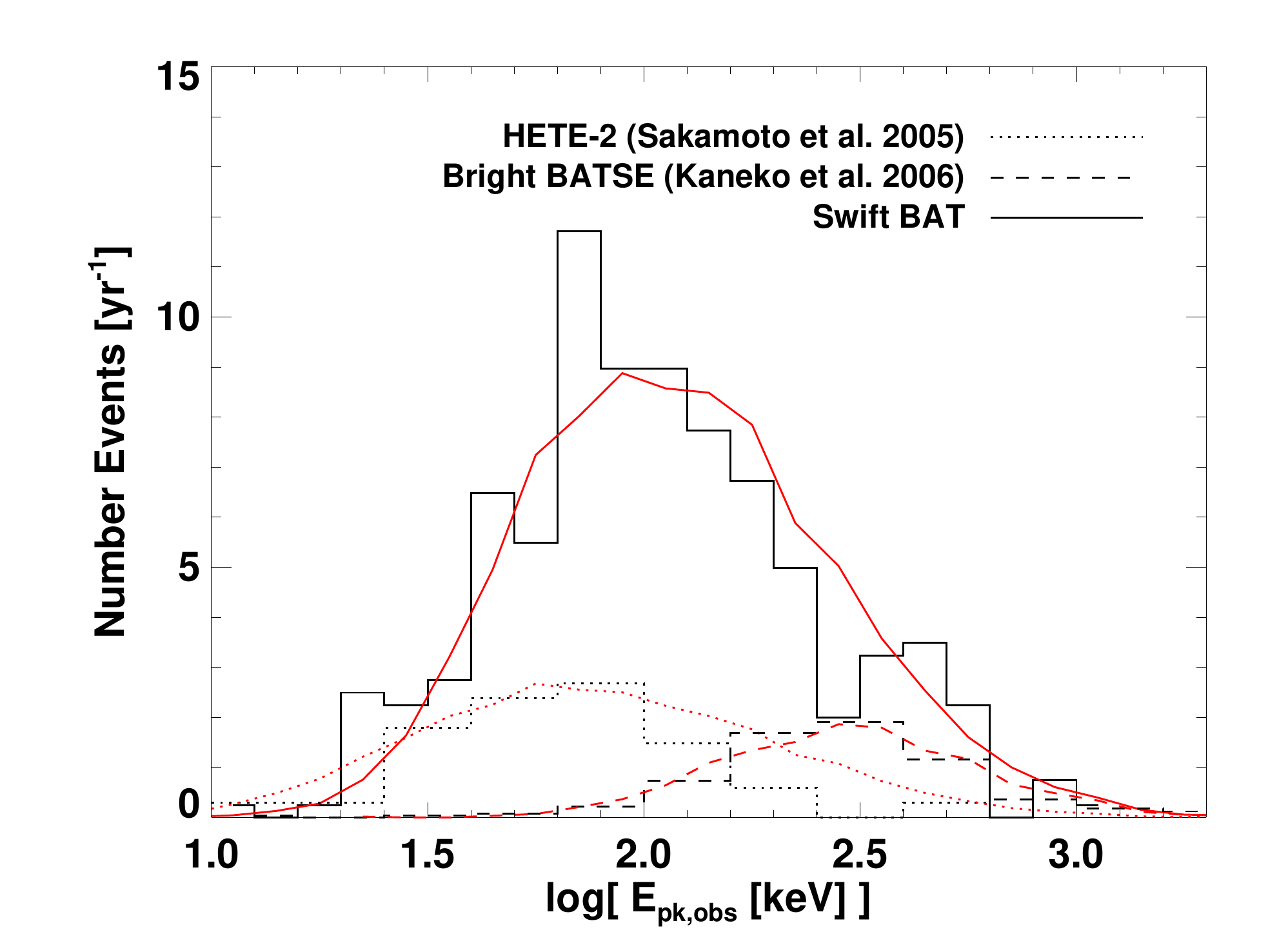}
\caption{\small 
The predicted $E_{\rm pk,obs}$ distributions for Swift (solid lines), BATSE (dashed lines), and HETE-2 (dotted lines)
agree well with the observed data (Section \ref{sec:dists}).  } 
\label{fig:ep_pred} 
\end{figure}

Although it was not a requirement of our fits,  a significant number of low-$E_{\rm pk,obs}$ events --- many of
which are not detected by Swift --- are required to account for the large relative number of X-ray Flashes \citep[XRFs;][]{heise01}
detected by HETE-2 \citep[e.g.,][]{taka05} and {\it Ginga}~\citep[e.g.,][]{stroh98}.  This has been discussed elsewhere, for example, by \citet{lamb05}.
Figure \ref{fig:ep_pred} shows the expected distributions in $E_{\rm pk,obs}$ for HETE-2 and BATSE.

The BATSE LAD trigger efficiency
relative to Swift BAT is taken from \citep{band06}.  To reproduce the relative fraction of bright GRBs in \citet{kaneko06}, we adopt
a factor 8 sensitivity decrease relative to the full BATSE sample.  From a KS-test ($P_{\rm KS}=0.32$), the predicted and observed distributions
do not differ strongly.  
    
The sensitivity curve for HETE-2 is taken to be the WXM sensitivity curve of \citet{band03}, with the modification
to account for $\Delta t>1$s triggers from \citet{lamb05}.  The WXM sensitivity is comparable to that of Swift BAT; however,
the total expected relative GRB rate is a factor $\approx 5$ times lower due to an unfortunate mismatch in the HETE-2 instrument fields-of-view \citep[see, e.g.,][]{preger01}.
Our resulting predicted $E_{\rm pk,obs}$ curve for HETE-2 is an acceptable fit ($P_{\rm KS}=0.38$)
to the observed data \citep[see,][]{taka05}.  In Section \ref{sec:exist} below, we further analyze the predicted $E_{\rm pk,obs}$ distribution in the
context of optimizing future GRB satellites to achieve maximal detection rates.

\begin{figure} 
\hspace{-0.2in}
\includegraphics[width=3.7in]{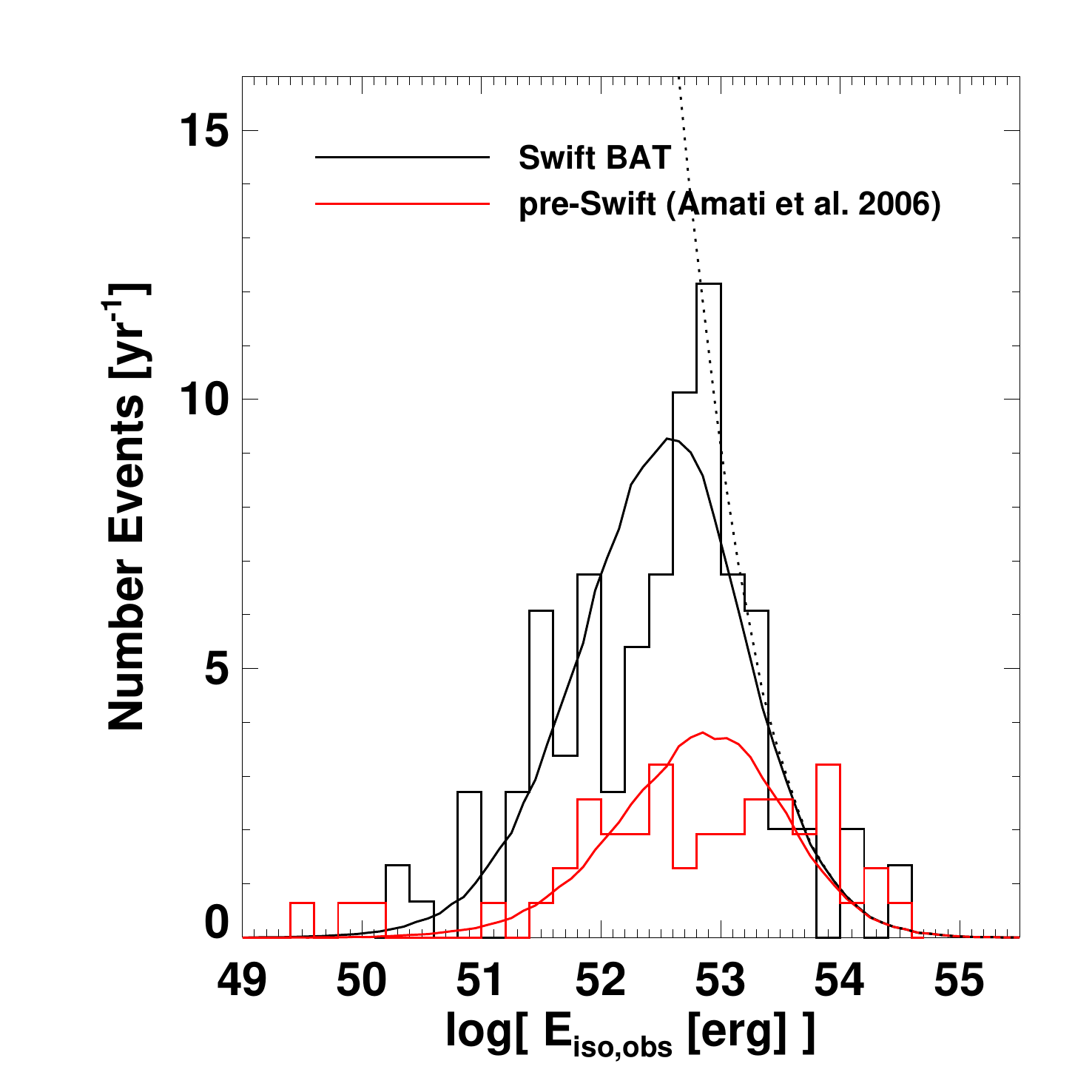}
\caption{\small 
The predicted $E_{\rm iso}$ distributions for Swift (black) and pre-Swift data
\citep[red; taken from][]{amati06} 
agree well with the observed data (Section \ref{sec:dists}).  The dotted line shows the distribution
without a detector cut-off: $\propto E_{\rm iso}^{a_L}$ below a smooth 
cutoff at $E_{\rm iso}\lessim 10^{53}$ erg.}
\label{fig:eiso_pred}
\end{figure}

To check for pre-Swift consistency in the predicted $E_{\rm iso}$ distribution (Figure \ref{fig:eiso_pred}) we take the $E_{\rm iso}$
sample from \citet{amati06}.  Completing this sample by accounting for the heterogeneous satellite thresholds 
is a challenging task to perform rigorously.  We assume, for simplicity, that the average threshold of pre-Swift satellites
tracks that of Swift but is a factor three times lower.  This translates to a total relative rate factor (Swift to pre-Swift)
of 2.5.  We find that the shape of the predicted distribution agrees well with the observed shape (KS-test $P_{\rm KS}=0.37$).

\begin{figure*} 
\includegraphics[width=4.0in]{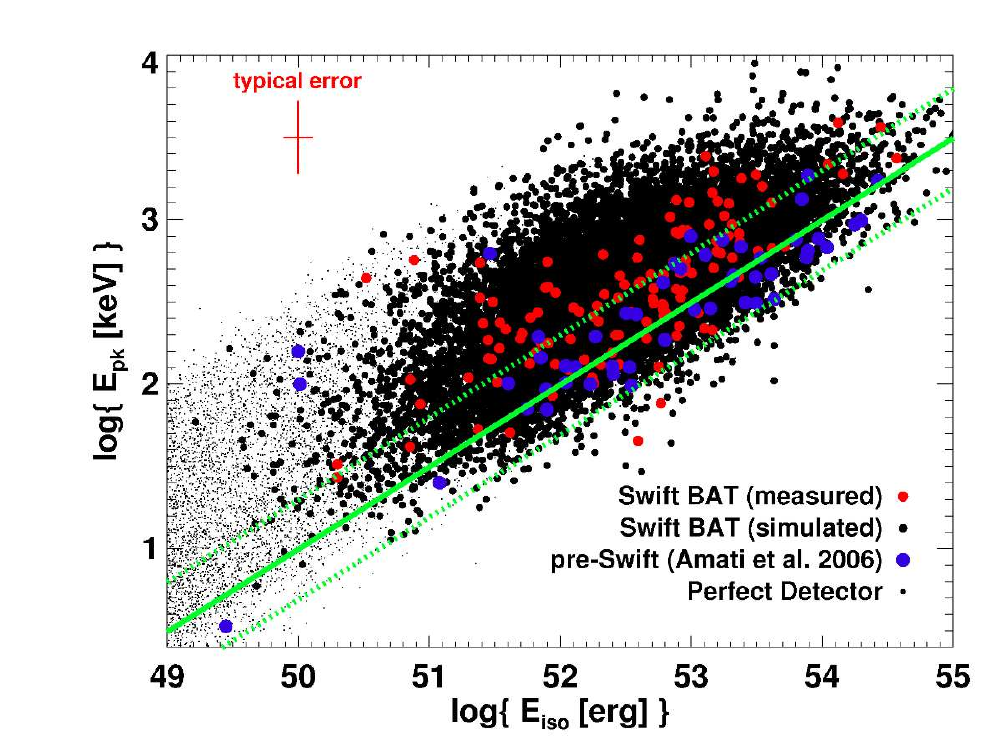}
\hspace{-0.2in}
\includegraphics[width=3.3in]{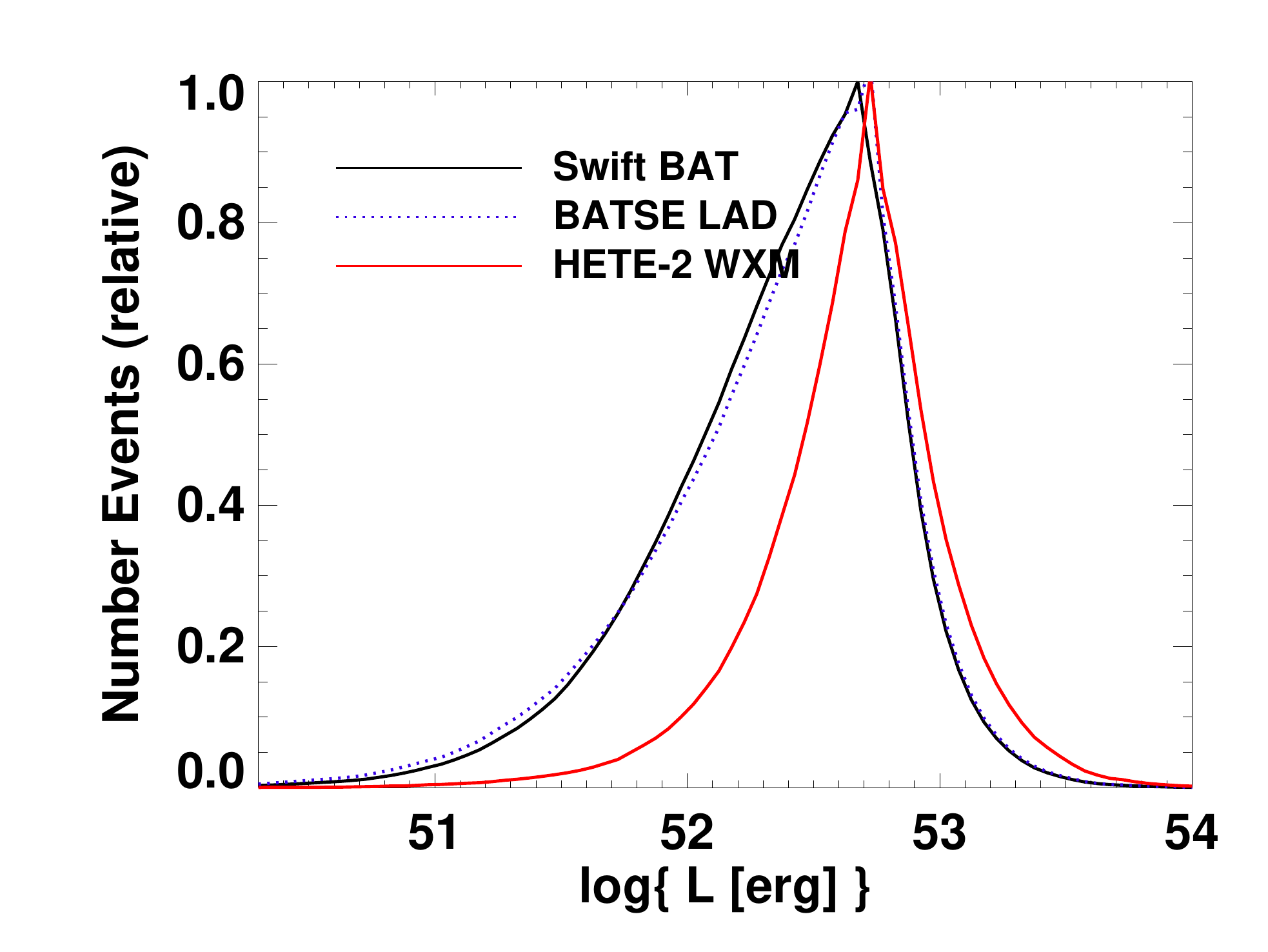}
\caption{\small 
The predicted $E_{\rm iso}-E_{\rm pk}$ distributions (A) and normalization (B) for multiple missions.
Due to the presence of faint GRBs which sample the low energy tail of the luminosity function (Section
\ref{sec:dists}, Swift and BATSE are expected to yield (A) a broad $E_{\rm iso}-E_{\rm pk}$ correlation \citep[as observed, e..g,][]{np05,bp05,butleretal07},
which increases in breadth as $E_{\rm pk}$ decreases.  Bright GRBs, from the least sensitive satellites, are expected to yield
a tight correlation \citep[as observed, e.g.,][]{amati02,amati06}.  The best-fit pre-Swift correlation and scatter are marked with green lines.  The normalization of the correlation (B) --- plotted most naturally here including.
duration and redshift (which contribute weakly to th observed shape) --- is shown to exhibit a strong decrease (factor 2)
in scatter in HETE-2 as opposed to Swift or BATSE.
A typical error bar for Swift is plotted in red in the upper left of sub-panel A.}
\label{fig:epeiso_pred}
\end{figure*}

In Paper I \citep[also,][]{butler09}, we study in detail the correlation in Swift between $E_{\rm iso}$ and $E_{\rm pk}$.
Compared to pre-Swift studies \citep[e.g.,][]{amati06}, a log--log fit to the correlation appears to exhibit a (factor 2) increase in scatter
and also a shift in normalization toward the Swift satellite threshold.  This suggests \citep[e.g.,][]{np05,bp05} the possibility that
the relation is actually an inequality: one region of
the $E_{\rm iso}-E_{\rm pk}$ plane (lower left in Figure \ref{fig:epeiso_pred}A) is physical and the other region is dominated by selection effects (upper
right in Figure \ref{fig:epeiso_pred}A).  A number of recent studies have examined the instrument-dependent population of faint GRBs \citep{ghirl08,nava08,sham09}.  The modelling here
sheds new light on this controversial area of study.

As we summarize above, we have parameterized an intrinsic correlation via a log--log slope parameter $\alpha_E$, which we find to be non-zero at a high
level of statistical significance.  The precise value of $\alpha_E$ depends sensitively on the rate of missing low-$E_{\rm pk,obs}$ GRBs (Section \ref{sec:fitting}).

It is important to stress that a $\alpha_E>0$ does not translate automatically into a strong observed correlation between
$E_{\rm iso}$ and $E_{\rm pk}$.  That requires also a narrow distribution in the effective luminosity $L$, which governs the observed slope and scatter.
Detected GRBs with $L \gtrsim L_{\rm cut}$, will exhibit a strong correlation.  However, if there are a significant number of GRBs extending to lower $L$ as a powerlaw in $L^{a_L}$, 
these will asymptotically not show a correlation.  Mathematically, this is because the luminosity function (Equation \ref{eq:lum_func}) becomes separable in $E_{\rm iso}$ and $E_{\rm pk}$.
A narrow log--log relation requires an effective luminosity function which is approximately a delta function.  More quantitatively, we find we must have $a_L> 2$
to generate a correlation with scatter $\lessim 0.15$ dex in the Swift data.   The Swift data rule out $a_L>2$ at the $6\sigma$ level ($-2\Delta \log(\mathcal{L})=40.2$ for one additional degree of freedom).

The expected behavior in the $E_{\rm eiso}-E_{\rm pk}$ plane is summarized in Figure \ref{fig:epeiso_pred}A, where we also compare to the Swift and pre-Swift data.  At high $E_{\rm iso}$, the relative
rarity of high-$E_{\rm pk}$ events leads to a strong correlation relatively independent of the detector.  However, at low-$E_{\rm pk}$ --- due to the luminosity function tail --- the 
correlation can exhibit a very large scatter.  Pre-Swift GRBs appear to occupy the ridge of events in Figure \ref{fig:epeiso_pred}A near $L=L_{\rm cut}$.  Swift has allowed us to sample well the broadening
of $L$ (see, also, Figure \ref{fig:fits}).  The BATSE faint GRB sample \citep{np05,bp05} also appears to behave this way (Figure \ref{fig:epeiso_pred}B).

The expected clustering of pre-Swift events near $L=L_{\rm cut}$ can be demonstrated by applying the instrumental thresholds as above from \citet{band03}
to our model.  As shown in Figure \ref{fig:epeiso_pred}B, the breadth of $L$ decreases as the satellite sensitivity decreases.  Also important is the relative lack of high-$E_{\rm pk,obs}$ sensitivity
for experiments like the WXM on HETE-2, which could permit the illusion that a true, narrow $E_{\rm iso}-E_{\rm pk}$ correlation extends to XRFs.  The FWHM of the distribution in 
$L$ is 0.4 dex for HETE-2 WXM as compared to 0.8 dex for Swift (and also for the faint BATSE sample).  

The HETE-2 WXM distribution in $L$ also has broad tails, which can make the observed distribution even narrower if tail events are falsely rejected as outliers.
The same effect may also be seen for the $E_{\rm iso}$ distribution in Figure \ref{fig:eiso_pred}, where a few subluminous events in the pre-Swift sample \citep[980425, 031203; e.g.,][]{sod04}
appear to form the tail in the distribution including Swift events.

We note that we find evidence for a possible correlation with duration $T_{r45,z}$ as in \citet{firmani06} for a considerably smaller sample \citep[but, see,][]{col08}.

\section{Discussion \& Results}

\subsection{Energetics vs. $z$}
\label{sec:energetics}

\begin{figure} 
\center{\includegraphics[width=3.5in]{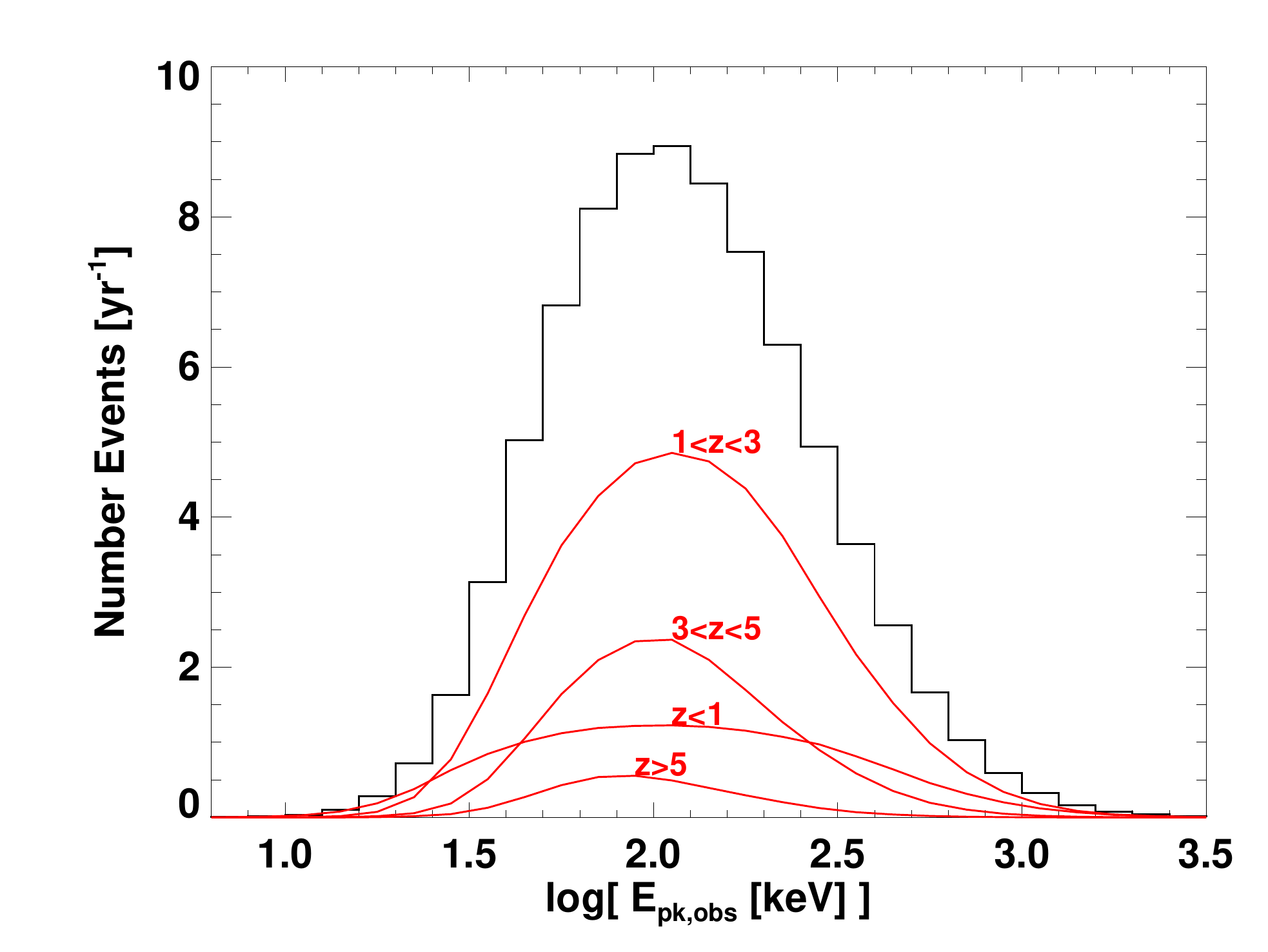}} 
\center{\includegraphics[width=3.5in]{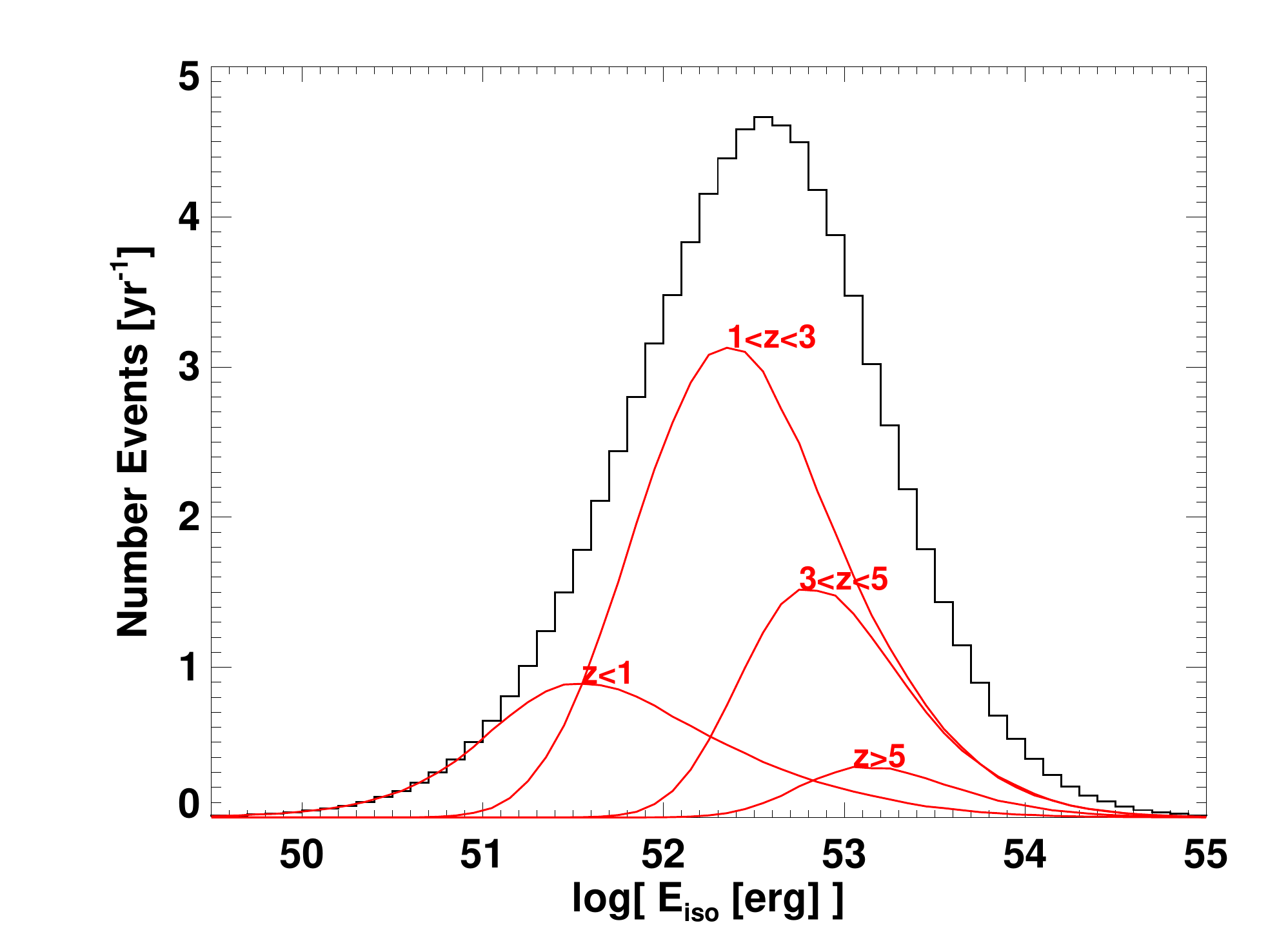}} 
\caption{\small 
The predicted variation of $E_{\rm iso}$ and $E_{\rm pk,obs}$ observed by Swift with redshift $z$ is weak.  High redshift (e.g., $z>5$) 
GRBs are systematically, intrinsically brighter than low$-z$ GRBs.  This is due primarily to the BAT flux limit, which corresponds to an
increasing $E_{\rm iso}$ with increasing $z$,
and not luminosity evolution, which is not significantly present in the Swift sample.} 
\label{fig:epeiso_z} 
\end{figure}

\begin{figure} 
\center{\includegraphics[width=3.5in]{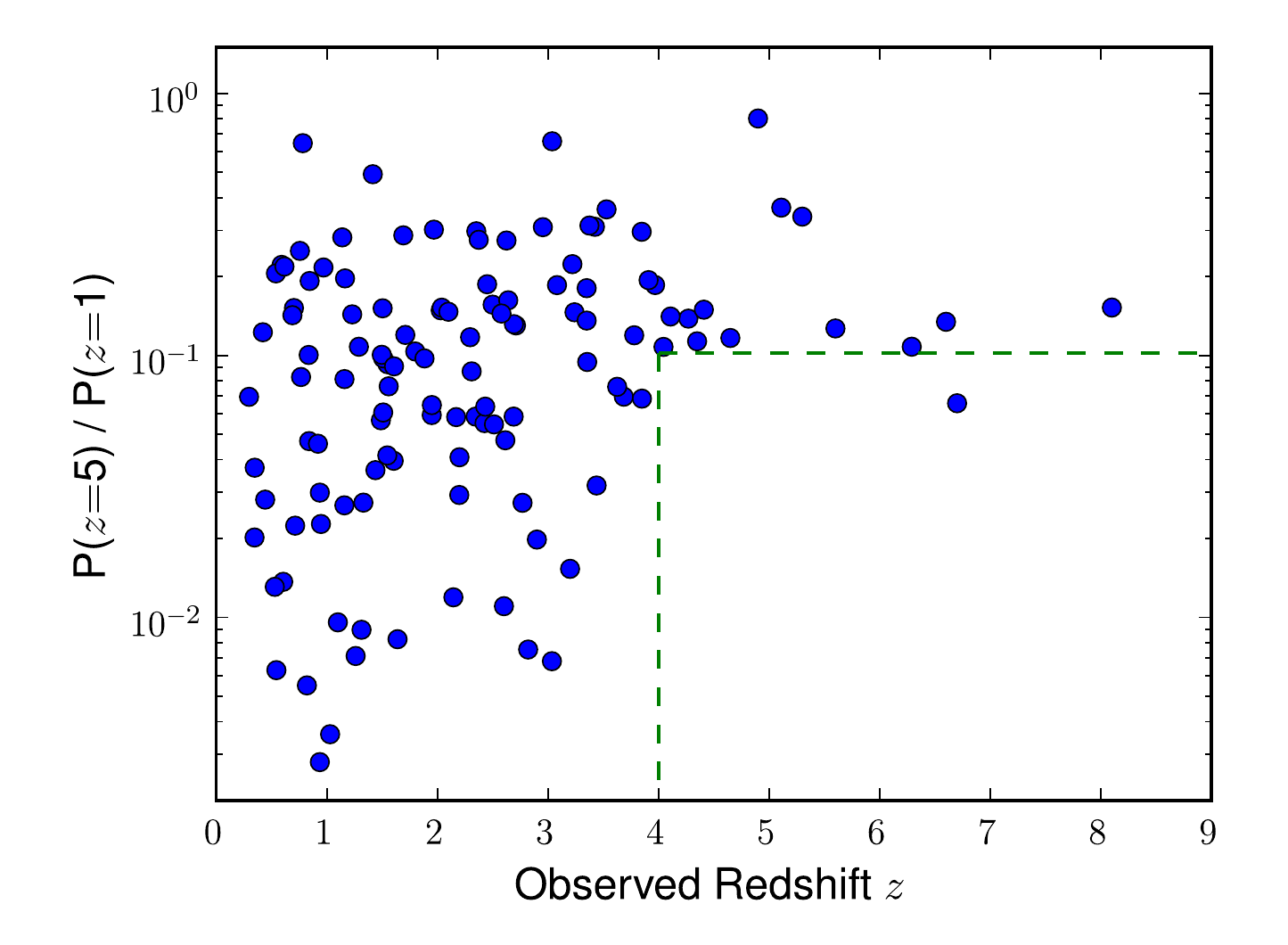}}
\caption{\small 
Ratio of the posterior probability at $z=5$ to that at $z=1$ for Swift bursts of known redshift $z$.  There is a large scatter
and only a weak correlation.  Weak decision rules can be obtained, however, resulting from the relative observer-frame faintness of high-$z$ GRBs above Swift threshold: nearly all (13/14) bursts at $z>4$ 
fall above the median P($z$$=$5)/P($z$$=$1) value for $z<4$ (dashed lines).  This allows a potential factor two
target reduction for high-$z$ followup.}
\label{fig:pred}
\end{figure}

We find a general tendency that high-$z$ GRBs are intrinsically brighter and
marginally softer in the observer frame than low-$z$ GRBs (Figure \ref{fig:epeiso_z}).
This arises predominantly due to the BAT flux limit, which corresponds to an increasing
luminosity with $z$.  Luminosity evolution (see below) does not play a strong role.
We note that the redshift dependence of the observables is weak.  
If we ask, for example, what is the posterior prediction for the redshift of a burst of known $z$,
we observe a large scatter between the predicted and true redshifts.  However,
weak decision rules can be obtained which may help identify high-$z$ GRBs (see Figure \ref{fig:pred}).

Contrary to some previous studies (Section \ref{sec:prior}), we do not find significant
evidence that the GRB luminosity function
evolves with redshift.  Although we have marginalized over
the possibility of redshift evolution in $E_{\rm iso}$,
the extent of that evolution is weak $E_{\rm iso} \propto
(1+z)^{0.0\pm0.5}$ (Table 1).

\begin{figure} 
\center{\includegraphics[width=3.5in]{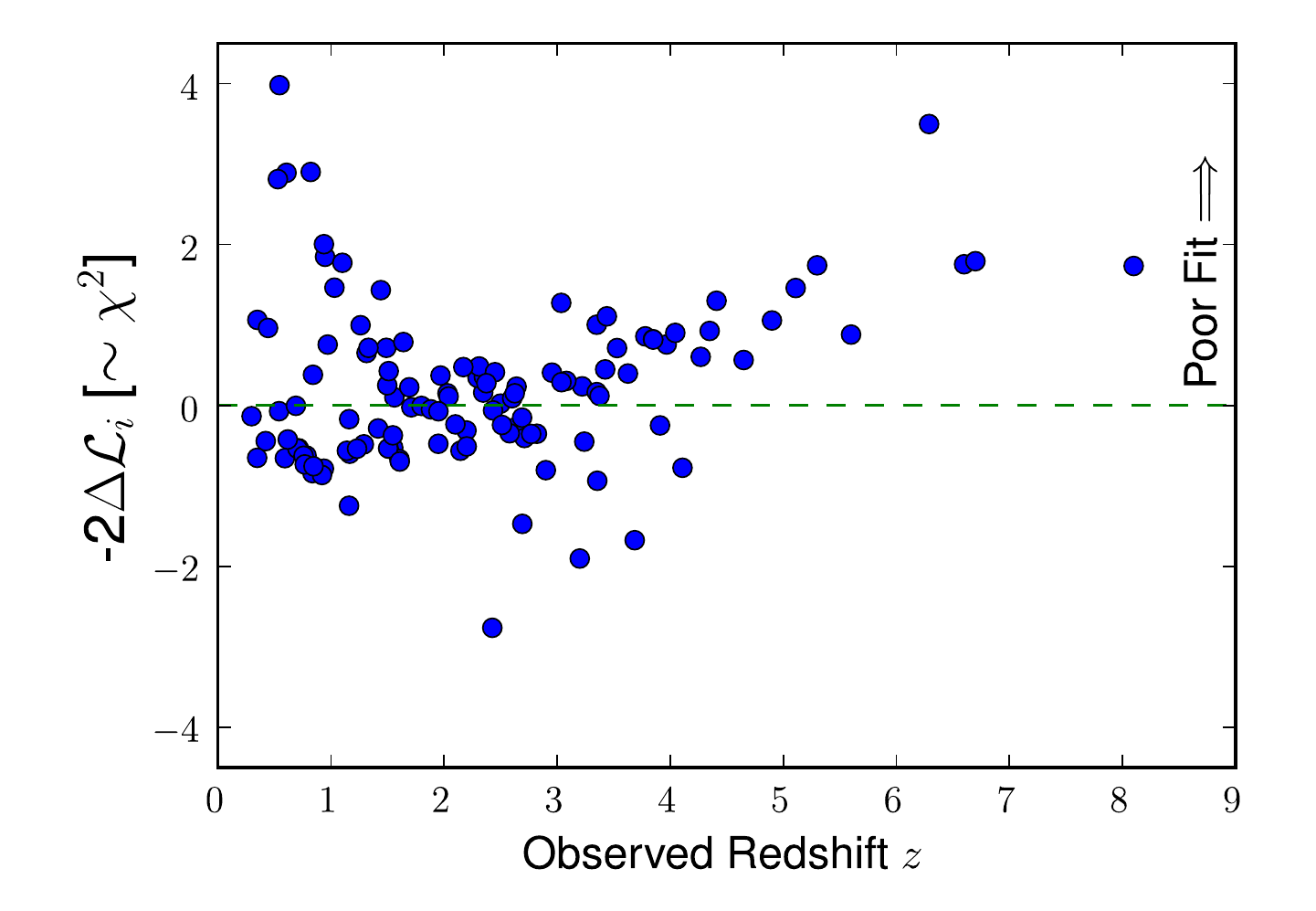}}
\caption{\small 
Decrease in the relative quality of fit $-2\Delta\log{(\mathcal{L}_i)}$ (Equation \ref{eq:logl}) for the best-fit model with
strong luminosity evolution ($\alpha_z \equiv 2$) relative to the best-fit overall model ($\alpha_z\approx0$; no evolution).  The $y$-axis scales approximately
as $\Delta \chi^2 \approx \sigma^2$ and is plotted for each GRB as a function of $z$.
The fit quality, considering
strong luminosity evolution, is systematically poor at low- and high-$z$, although there are a few GRBs at intermediate
$z$ which appear to prefer evolution. The overall evidence in disfavor of strong luminosity evolution is $-2\sum \Delta\log{(\mathcal{L}_i)} = 25.1$
($5\sigma$; Section \ref{sec:energetics}).}
\label{fig:evol_fit}
\end{figure}

We can test explicitly
for strong luminosity evolution ($\alpha_z\geq 2$).  We obtain
fits which have a lower GRB rate density at intermediate and high redshift ($g_1 \approx -0.1$
instead of $g_1>1$ and $g_2 \approx -5$ instead of $g_2\approx -3$ 
for $\alpha_z \approx 0$, with only modest change in the other model 
 parameters; see also Section \ref{sec:fitting}).
However, the decreased quality of the fits
($-2\Delta\mathcal{L} = 25.1$ for 1 additional degree of freedom) ---
resulting from a relatively poorer fit to sample bright GRBs at low-$z$ and faint GRBs at high-$z$ ---
rules out $\alpha_z\geq 2$ at the $5\sigma$ level (see, Figure \ref{fig:evol_fit}).
To favor strong luminosity evolution, we would need to
assume strong prior information that the GRB rate density follows more 
closely
the cosmic star formation rate (see, Section \ref{sec:rho}).

\subsection{The GRB Comoving Rate Density}
\label{sec:rho}

\begin{figure*} 
\center{\includegraphics[width=6.5in]{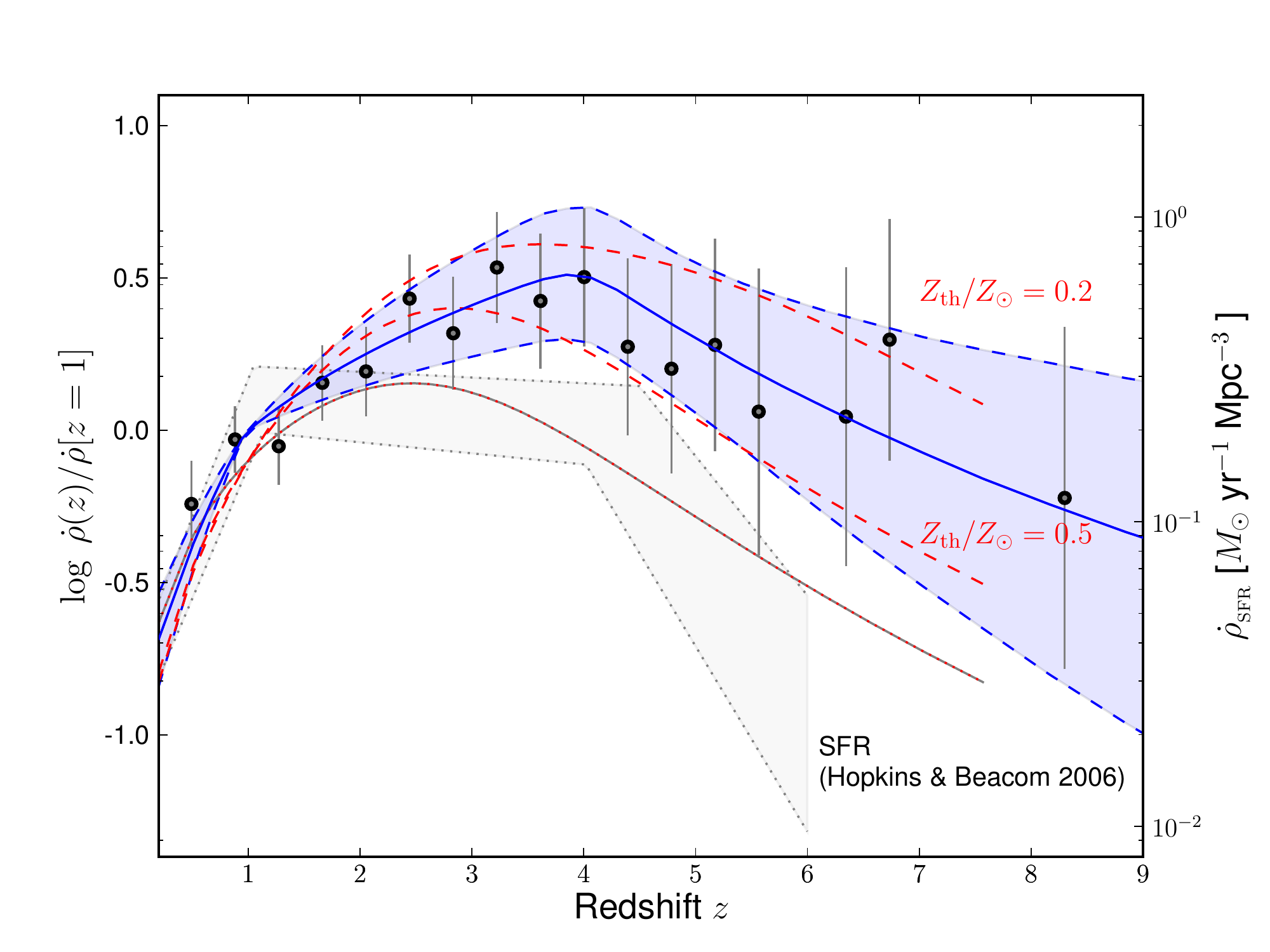}}
\caption{\small 
The comoving rate density, relative to the rate density at $z=1$.  Dashed lines give the 90\% confidence regions for the best fit (blue) curve.
Also plotted is the location of star formation rate (SFR) data (gray hatched region) and a 
parameterized fit (solid gray curve) from \citep{hopkinsbeakom}.  We use this fact that the GRB rate traces well the SFR at $z<1$ to normalize the GRB 
rate (right $y$-axis).  Following \citep{salv07}, the red dashed curves show plausible enhancements to the gray SFR curve
resulting from a metallicity cutoff: GRB hosts have $Z<Z_{\rm th}$.}
\label{fig:rate_dens} 
\end{figure*}

The best-fit GRB rate density (Equation \ref{eq:rho}) --- marginalized over GRB luminosity evolution, etc. --- and uncertainty are shown in
Figure \ref{fig:rate_dens}.  We find that the $z\lessim 1$ slope $g_0$
appears to track that of star formation well.  This fact can be used to approximately
normalize the relation at $z=1$.  A rigorous normalization is not possible here, because the Swift
data do not constrain the number of GRBs at the faint end of the luminosity function nor is the beaming
fraction $\langle f_{\rm beam} \rangle$ well-constrained.
We can also quote a normalization above a given intrinsic flux level:
the all-sky intrinsic rate of long-duration GRBs with $E_{\rm iso}>10^{52}$ erg is  $10^{3.1\pm0.3}$ GRBs $\langle f_{\rm beam} \rangle^{-1}$ yr$^{-1}$
(or $10^{-2.4\pm0.5}$ GRBs $\langle f_{\rm beam} \rangle^{-1}$ yr$^{-1}$ Gpc$^{-3}$ at $z=0$).
Integrating the SFR over volume and
considering a Salpeter IMF \citep[e.g.,][]{dt99} with GRB progenitors more massive than 20 $M_{\odot}$ \citep[e.g.,][]{fryer99}, only about 0.1\% $[\langle f_{\rm beam} \rangle/0.01]$ of
massive stars explode as bright GRBs \citep[see, also, e.g.,][]{sod05}.  Numbers increase mildy to low $E_{\rm iso}$ levels $\propto E_{\rm iso}^{-0.36\pm 0.07}$.

Beyond $z=1$, the GRB rate increases significantly faster than star formation.
At $z=6$ the GRB rate is roughly two orders of magnitude greater,
although the star formation rate is quite uncertain at such high redshift.
The GRB rate enhancement relative to star formation, which we derive for $z=1-4$, is closely consistent with that found 
by \citet{kistler08} using a smaller Swift sample
and not fitting for the possibility of GRB luminosity evolution.

\citet{kistler08} discuss a number of possible scenarios which
could yield that evolution (decreasing cosmic metallicity with $z$, an initial mass function for stars which becomes increasingly skewed
toward more massive stars at high-$z$, etc.), but conclude that no scenario adequately describes the data without fine-tuning.
\citet{salv07} \citep[also,][]{modjaz08,salv09a,salv09b} have explored one such fine-tuning scenario --- a preference for GRBs to arise in metal-poor host 
galaxies \citep[e.g.,][]{savag06,stan06},
resulting in a metallicity cutoff $Z<Z_{\rm th} \approx 0.1 Z_{\odot}$.
Following \citet{salv07} \citep[also,][]{ln06}, we can use the prescription from \citet{kewl05} for how metallicity evolves with redshift to estimate 
the mass density $\Sigma$ evolution:
\begin{equation}
\Sigma(z) = { \hat \Gamma[ 0.84, (Z_{\rm th}/Z_{\odot})^2 10^{0.3z} ] \over \Gamma(0.84) },
\end{equation}
where $\hat \Gamma$ ($\Gamma$) is the incomplete (complete) gamma function, $\dot \rho(z) \propto \Sigma(z) \dot \rho(z)_{\rm SFR}$.
This translates into a $\dot \rho(z)$ that peaks at higher $z$ than $\dot \rho_{\rm SFR}$.

\citet{salv07} \citep[also,][]{salv09a} decide that luminosity evolution (a possibility we rule out at the $5\sigma$ level; Section 
\ref{sec:energetics}) fits the data better than a strong metallicity cutoff
$Z_{\rm th}/Z_{\odot}=0.1$.  However,
we find that a more relaxed cutoff ($Z_{\rm th}/Z_{\odot}$ in the range 0.2---0.5; Figure \ref{fig:rate_dens}) appears to describe the data quite well.
A higher cutoff is also more consistent with studies of the GRB host galaxy mass distribution \citep[e.g.,][]{koc09}.
The apparently smooth continuation of the rate density above $z=4$ suggests the presence of no new evolutionary effects
\citep[see, also,][]{kistler09}.  The uncertainty is large, however.  It is possible that Pop. III stars may begin to contribute at this epoch,
although, our observed rates (Section \ref{sec:exist} below)  are marginally inconsistent with the prediction from \citet{bnl02} of 10\% of Swift GRBs at $z>5$.

\subsection{The $z$ Distribution for GRBs Without $z$}
\label{sec:redshift}

We have ignored in the above analysis the potential selection effects associated with obtaining spectroscopic host redshifts \citep[e.g.,][]{bloom03}.
In principle,
relative to pre-Swift experiments, arcsecond X-ray positions from the XRT for nearly all GRBs \citep[e.g.,][]{butler07} should translate into a higher optical followup and detection
rate. However, it remains true in the Swift era that 25--50\% of all GRBs exhibit suppressed optical flux relative to X-ray flux or are undetected
\citep[``dark'' GRBs; e.g.,][]{jak04,melandri08,cenko09,zheng09}
despite deep followup.  We have redshifts for only 37\% of Swift GRBs.  The distribution of these redshifts depends, in principle, not just on the intrinsic distribution but also 
on the complicated details of GRB afterglow observability \citep[location in the sky, time of month, etc.; see, e.g.,][]{jak05} and also on the
relative interest of particular observers to expend valuable resources on GRBs lying at a given potential redshift.
We explore here the biases these effects could introduce.

We have conducted a number of 2-sample KS tests to check consistency between the distributions of parameters in Tables
3 \& 4 for GRBs with $T_{90}>3$ s.  The duration statistics show no signs of 
strong disagreement ($P>0.7$).
There is modest evidence ($P=0.1$) for difference in the 15--350 keV fluences (also present 
for the bolometric fluence and $E_{\rm pk,obs}$).  We do note that the events without $z$ are on 
average about 50\% fainter.  
These events could be fainter because they, on average, have lower intrinsic luminosity
or because they are at higher $z$ (or a combination of the two).  
That such a large fraction events shows this flux variation suggests a range of source 
redshifts and the probability that most correspond to $z\approx 1-3$ where the observable Universe has most of its volume.

We can investigate a spectroscopic redshift selection bias by fitting for it explicitly.
Here, we imagine that $\dot \rho(z)_{{\rm no-}z} = \dot \rho(z)$, the true
rate density, while
$\dot \rho(z)_{z} = \epsilon(z) \dot \rho(z)$, where $\epsilon(z)$ is a function describing the optical detection and spectroscopy rate of GRBs and their host galaxies versus $z$.
We allow $\dot \rho(z)_{{\rm no-}z}$ and $\dot \rho(z)_{z}$ to have different 
slopes $g_1$ and $g_2$ describing the intermediate and high-$z$ populations.  It is reasonable to expect spectroscopy to achieve high recovery rates ($\epsilon(z)$ not decreasing)
below $z=1$ \citep[e.g.,][]{bloom03}.

We find that these extra two degrees of freedom in the modelling allow for a meager 1$\sigma$ improvement
in the fit ($-2\Delta\mathcal{L}=2.74$).  
Similar to the argument above based on concordance of fit parameters, this tells us that the sample without $z$ does not strongly demand a different $\dot \rho(z)$.
However, to what extent {\it could}~the distributions differ and how does this impact our conclusions?
The best-fit slopes for $\dot \rho(z)_{{\rm no-}z}$ ($g_1=0.75$, $g_2=-3.6$) are
modestly lower than those for $\dot \rho(z)_{z}$ ($g_1=1.2$,$g_2=-2.7$), indicating a bias $\epsilon(z) \sim (1+z)^{0.5}$ against obtaining redshifts for low-, rather than high-$z$ GRBs.
Marginalizing over the extra parameters, we find that $\dot \rho(z)$ is affected at the $<10$\% level for $z=1-4$.  Therefore, our constraints above on density evolution are practically unaffected.

The bias has a modest effect on the predicted rates of high redshift GRBs below:
$(30\pm30)$\% {\it fewer}~$z>6$ GRBs are expected in the full Swift (or EXIST) sample as compared to assuming that $\epsilon(z)$ is independent of $z$.  
The extra uncertainty in the rates resulting from marginalizing over $\epsilon$ is small compared the uncertainty already present at high-$z$.
The reader should be warned, however, that --- due to the small sample size at high-$z$ --- we cannot be sure that our parameterization of $\epsilon$ is sufficiently robust to reliably account for 
high-$z$ rate variations.

The most natural explanation for --- or at least a strong contributor to --- the apparent lack of low spectroscopic redshifts may be the so called 
``redshift desert'' at $1.5\lessim z \lessim 2$ \citep[the observed rate does in fact show a
decrement here; Figure \ref{fig:fits}; see also,][]{coward08}. There is a lack of strong star-formation emission lines in the optical bands for galaxies observed in this redshift range.
It is possible to obtain redshift in this range (detected, e.g., through Mg~II absorption of bright afterglow light), but host galaxy redshifts will be very challenging to
obtain for faint afterglows or those observed too late.
It is quite likely that some or many Swift GRBs without spectroscopic redshifts have true redshifts preferentially in the range $1.5<z<2$.  Deep host galaxy 
imaging studies \citep[e.g.,][]{perley09,fynbo09}
can shed important light on this possibility, because these hosts may in general be detectable with modest imaging investments.

\begin{table}[htbp]
\begin{center}
\caption{Predicted Redshift Rates [yr$^{-1}$]}
\begin{tabular}{l|c|c}\hline\hline
         &    Swift          &   EXIST \\\hline
all$-z$  &  89 $\pm$ 8   & 270 $\pm$ 40 \\
$5<z<8$  &   4.4 $\pm$ 2.4 & 17.4 ($-7.6$,$+8.5$) \\
$8<z<12$ &   0.5 ($-0.5$,$+0.9$) & 2.3 ($-2.1$,$+3.7$) \\
 & & \\
$z>5$ & 5.0 ($-2.8$,$+3.3$) & 20.2 ($-9.6$,$+13.5$) \\
$z>6$ & 2.3 ($-1.8$,$+2.3$) & 9.6 ($-6.2$,$+10.1$) \\
$z>7$ & 1.1 ($-1.1$,$+1.6$) & 5.0 ($-3.9$,$+7.3$) \\
$z>8$ & 0.6 ($-0.6$,$+1.1$) & 2.8 ($-2.5$,$+5.3$) \\
$z>10$ & 0.2 ($-0.2$,$+0.6$) & 1.0 ($-1.0$,$+2.9$) \\
$z>12$ & 0.1 ($-0.1$,$+0.4$) & 0.5 ($-0.5$,$+1.7$) \\\hline
\end{tabular}
\end{center}
\label{tab:rates}
{Note.---Rates (90\% Conf.) for long duration $T_{90}>3$ s GRBs only.  Short-duration GRB rates are discussed in Section \ref{sec:exist}.}
\end{table}

\subsection{Towards a Better Mousetrap: EXIST}
\label{sec:exist}

In this section, we utilize the best-fit Swift GRB model (Table 1, left)
to estimate the number of GRBs Swift --- and a more sensitive future Swift-like experiment ---
will detect as a function of $z$.
EXIST \citep[e.g.,][]{grindlay09},
to a reasonable approximation, with BAT-like CZT detectors and trigger software,
is a scaled-up (i.e., more sensitive)
Swift BAT \citep[see,e.g.,][]{band03,band08}.  We take the sensitivity increase to be a factor 7.2 and the field of view to be
the same at the BAT (Josh Grindlay, private communication).

\begin{figure} 
\includegraphics[width=3.7in]{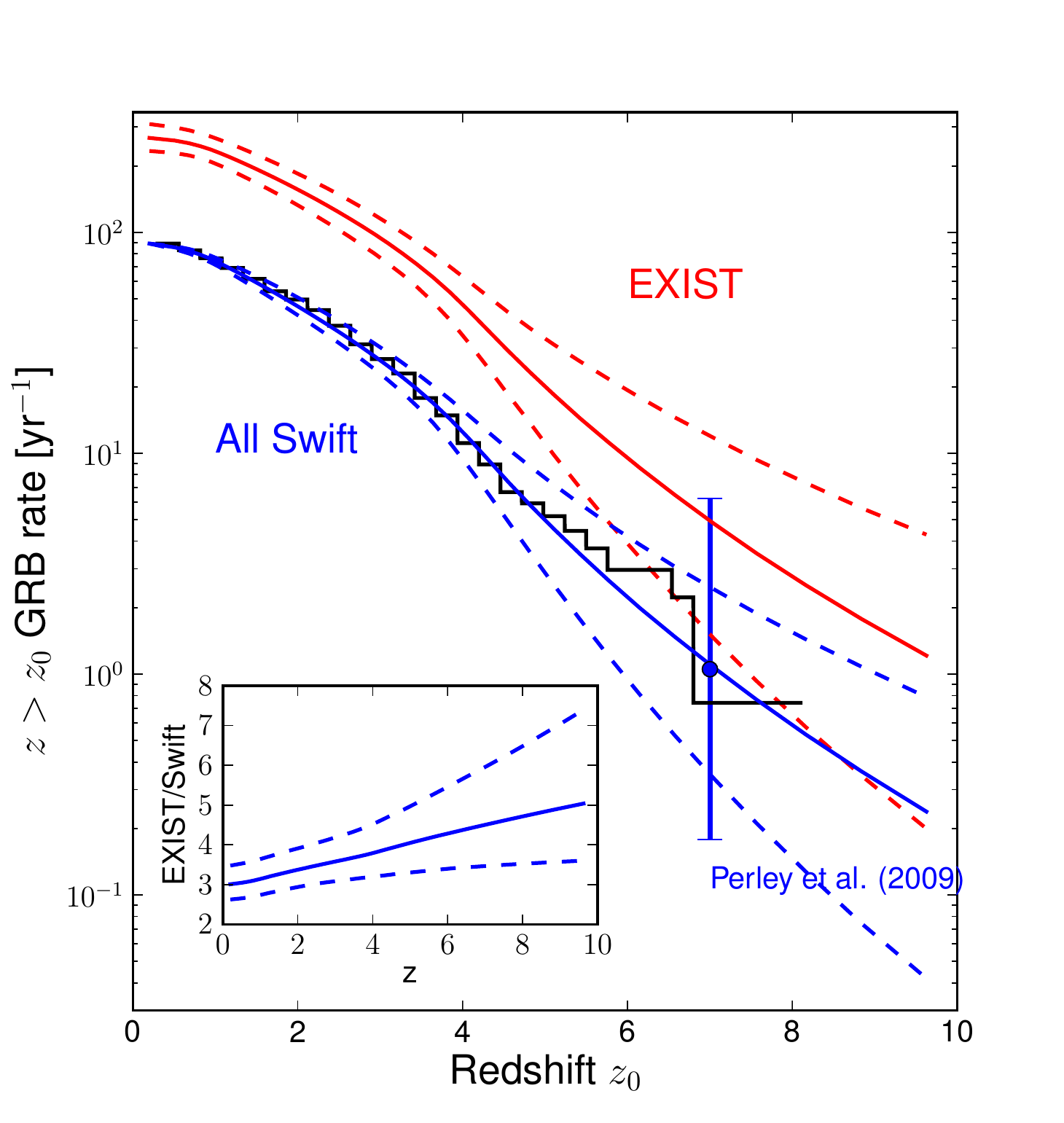} 
\caption{\small 
The observed (black) and predicted redshift rates for Swift (blue) and EXIST 
(red) GRBs.  The 90\% confidence intervals are marked with dashed lines 
(see, also, Table 2).  The plot subpanel demonstrates the increasing, relative 
sensitivity of EXIST at high $z$ by showing the relative 
number of predicted GRBs above redshift $z$.   The blue point from \citet{perley09} is an independent measurement based on optical studies of optically-dark  GRB
host galaxies.} 
\label{fig:exist} 
\end{figure}

Figure \ref{fig:exist} shows the expected rate of detectable long-duration ($T_{90}>3$ s) GRBs for Swift --- if redshifts were
measured for all Swift GRBs --- and for EXIST (see, also, Table 2).  About 3--9\% (4--13\%) of all Swift (EXIST) GRBs are 
expected to lie at high redshift ($z>5$).  Our prediction is consistent with, but on the low side of, that found by
\citet{jak05,jak06} (5--40\% of Swift GRBs at $z>5$) from a survival analysis of a sub-sample of Swift GRBs with uniform optical followup
properties and redshifts (or redshift constraints).
Our numbers agree well with those presented in \citet{salv08}, provided we correct the
EXIST field-of-view from \citet{salv08} to be 1.4 sr instead of 5 sr.  Our numbers also agree well with independent estimates
based on host-galaxy observations of optically dark GRBs \citep[0.2--14\% \{0.2--7\%\} of Swift GRBs at $z>5$ \{$z>7$\}][see Figure \ref{fig:exist}]{perley09}.

The relative EXIST/Swift rate increases with increasing redshift (subpanel Figure \ref{fig:exist}).
This can be understood as the result of the $E_{\rm iso}-E_{\rm pk}$ correlation and the 
redshift dependence of the flux (Figure \ref{fig:epeiso_z}) --- a more sensitive satellite detects fainter, softer GRBs, and these tend
to be preferentially at higher redshift.  The strong difference in expected rate-increase-versus-hardness is most clearly seen in a plot
of the Swift logN--logS (e.g., Figure \ref{fig:logN-logS}).  We observe that faint, hard GRBs are becoming increasingly rare, while faint, soft
GRBs are growing strongly in number.  It is also interesting to note (see, Figure \ref{fig:logN-logS}) that the short GRB rate appears to
be steeply rising to low flux levels.  This suggests that EXIST may see a factor ten more short GRBs ($\gtrsim 120$ yr$^{-1}$) than Swift, extending
the observed short-duration GRB redshift sample to intermediate and possibly high-$z$.

We now explore the characteristics of an optimal GRB detector, so constructed as to maximize the number of
detected GRBs.
First, it is important to note that a large sensitivity increase (factor 7.2 for EXIST relative to Swift) results in only a factor $\approx 3-4$
increase in rates due to the shallowing out of the logN--logS (Figure \ref{fig:logN-logS}).  Therefore, field-of-view increase should be prioritized
over sensitivity increase.  Second, it is clearly important to be able to trigger over a broad range of timescales \citep[e.g.,][]{band06}.
Finally, the analysis above allows us to make general recommendations regarding spectral sensitivity.
The observed peak in $E_{\rm pk,obs,max}$ in the $E_{\rm pk,obs}$
distribution is at $\approx 100$ keV for Swift.
If we imagine a hypothetical detector which is $S_{\rm rel}$ times
more sensitive than Swift,
then we observe that the peak in the $E_{\rm pk,obs}$ distribution decreases 
as $E_{\rm pk,obs,max} \propto S_{\rm rel}^{-1/4}$.  Consistently, as the peak decreases, 90\% of 
all GRBs have $E_{\rm pk,obs}<E_{\rm pk,max}/3$.
The rate of GRBs with $E_{pk,obs}=10$ keV relative to those
with $E_{\rm pk,obs}=100$ keV, for example, increases strongly as 
$\approx S_{\rm rel}^2$.
That is, a doubling of $S_{\rm rel}$ results in about three times more
$E_{\rm pk,obs}=10$ keV bursts and only about 30\% more $E_{\rm pk,obs}=100$ keV
GRBs (see also, Figure \ref{fig:logN-logS}).
It does not help to increase sensitivity to hard GRBs, because these
are rare.  Optimizing for high GRB rates means optimizing for soft GRBs.
A satellite which detects soft GRBs well is also an efficient high-$z$ machine, because 
high-$z$ GRBs have lower than averge $E_{\rm pk,obs}$ (Figure \ref{fig:epeiso_z}).

\section{Conclusions}
\label{sec:concl}

We have exploited a careful modelling of the Swift BAT trigger 
threshold --- strongly dependent on burst
hardness and duration --- to derive constraints on the GRB intrinsic distributions in $E_{\rm iso}$, $E_{\rm pk}$,
$T_{r45,z}$, and $z$.
This is the first study, utilizing a uniform reduction of the 
largest available number of Swift GRBs, to rigorously propagate
GRB measurement errors and their covariances
(through very general functional forms describing the intrinsic 
distributions) to 
obtain reliable confidence intervals for the rates.  The 
modelling is now possible thanks to the large number of spectroscopic 
redshifts
for Swift GRBs which largely alleviate parameter degeneracies.

We find that the GRB luminosity function is best written as a broken powerlaw (Table 1) in the effective luminosity 
\begin{equation}
L = {E_{\rm iso} \over (1+z)^{0.0\pm0.5}} \left({10^{2.5} {\rm keV} \over E_{\rm pk}} \right)^{1.8\pm0.3} \left( {10^{0.6} {\rm s}\over T_{r45,z}} \right)^{0.4\pm0.2}.
\end{equation}
The distribution $\tilde \phi(L) = dN/d\log{L}$ is nearly flat $\propto L^{-0.2\pm0.2}$ below $\log{(L_{\rm cut})} = 52.7\pm0.4$ and declines sharply thereafter
$\propto L^{-3.0\pm1.5}$.
Our multivariate GRB world model, which includes this luminosity function to produce very different intrinsic rates 
for hard or soft-spectrum GRBs (as observed, e.g., Figure \ref{fig:logN-logS}),
  accurately reproduces both Swift and pre-Swift rates, and should be very useful
for future simulation studies.
We note that the overall rate versus redshift does not appear to strongly depend on the multivariate modelling.

We draw the following principle conclusions:
\begin{itemize}

\item There is a real, intrinsic correlation between $E_{\rm iso}$ and $E_{\rm pk}$ (and possibly also
$T_{r45,z}$), despite indications in our previous work (Paper I) that this might be solely due to a selection effect.  However, the correlation {\it is not} a narrow log-log relation \citep[e.g.,][]{amati08}, and its observed appearance (more like an inequality) is strongly detector-dependent.

\item We find modest ($\sim$$3\sigma$) evidence for a large population of low-$E_{\rm pk,obs}$ GRBs \citep[XRFs; see, also][]{stroh98,lamb05}, and future satellites may overwhelmingly detect these, as well as a potentially large missing (from Swift) population of short-duration GRBs.

\item The Swift sample does not require luminosity evolution to produce the observed number of GRBs at high-$z$.  In fact, we rule out the possibility of strong evolution at the $5\sigma$ level.

\item An increase in the GRB rate relative to the SFR at $z>1$ appears to be due evolution in the rate density, and the shape of this evolution is roughly consistent with that expected from a preference for GRBs to reside in modestly low-metallicity ($Z \lessim 0.3 Z_{\odot}$) environments.

\item Finally, the predicted observed rate of GRBs at high-$z$ ($3-9$\% at $z>5$) is consistent with, if marginally lower than that found in previous studies.
\end{itemize}

The conclusions above are based primarily on the high-energy properties of Swift GRBs, without detailed consideration
of the afterglow properties or optical selection biases (but see Section \ref{sec:redshift}).  This is good in the sense that our rate
estimates will be largely independent of those based primarily on afterglow or host galaxy studies \citep[e.g.,][]{perley09,fynbo09}.
However, we are fundamentally limited in firmly extending our conclusions to include all GRBs
by the fact that a majority of Swift GRBs lack measured redshifts.
We note that redshift limits are easily incorporated into
the formalism we have developed; however, our initial calculations suggest there are currently too few \citep[about
20\% for dark GRBs][]{jak05,jak06} to significantly improve or modify our results.
Also, useful would be constraints on the highest energy emission from simultaneous observations from
other satellites \citep[e.g.,][]{bellm,krimm}.  These would help limit the sharpness of the luminosity
function cutoff $L_{\rm cut}$ to help infer the breadth of the intrinsic luminosity distribution and also
better limit the potential of luminosity evolution.  Finally, it will be important to better constrain the 
role of
GRB beaming and the way this shapes the distributions in numbers, $E_{\rm iso}$, $E_{\rm pk}$, etc.
\citep[e.g.,][]{guetta05,lamb05}.
Given apparent challenges in inferring beaming from imaging observations of X-ray and optical afterglows
\citep[e.g.,][and references therein]{racusin09}, it may be most fruitful to focus on late-time radio observations \citep[e.g.,][]{cenko09c}.

\acknowledgments
We thank D. Perley, D. Kocevksi, B. Cenko, M. Kistler, and N. Gehrels for comments on the manuscript and for useful discussions.
Greatly appreciated also were helpful comments and criticisms from an anonymous referee.
NRB is supported through the GLAST Fellowship Program (NASA Cooperative Agreement: NNG06DO90A).
D.P. is partially supported by US Department of Energy SciDAC grant DE-FC02-06ER41453.

\appendix
\renewcommand{\theequation}{A-\arabic{equation}}

To rigorously accounts for measurement errors and for correlations present in the data,
we characterize the GRB rate as a product of terms describing the intrinsic distributions in $z$, $E_{\rm iso}$, $T_{r45,z}$, and $E_{\rm pk,obs}$.
Expanding the notation from Equation \ref{eq:rtrue}; Section \ref{sec:redux}, the
the true, detector-independent event $N$ differential rate is:
\begin{equation}
r_{\rm true} = {dN \over d\log[E_{\rm iso}] d\log[E_{\rm pk}] d\log[T_{r45,z}] dz} =
P_{\rm corr}(E_{\rm iso}| E_{\rm pk},T_{r45,z},z ) P_E(E_{\rm pk}) P_T(T_{r45,z}) {r_0 \dot \rho(z) dV/dz \over (1+z)},
\end{equation}
where $P_{\rm corr}$ is a luminosity function that allows for the potential dependence of $E_{\rm iso}$ on hardness, duration, and redshift.

We consider a regression model for $P_{\rm corr}$:
\begin{equation}
P_{\rm corr} = \int d\log{[E_0]} { \phi_L(E_0) \over \sqrt{2\pi\sigma_L^2}}
\exp \bigl( -0.5 [ \log{(E_{\rm iso})} - \log{(E_0 E_{\rm pk}^{\alpha_E} T_{r45,z}^{\alpha_T} (1+z)^{\alpha_z})} ]^2/\sigma_L^2 \bigr),
\end{equation}
where $\phi_L$ describes the normalization $E_0$ of the intrinsic correlation between luminosity, duration, hardness, and redshift.  We can carry-out the integration, which can be regarded as a Gaussian smoothing of $\phi_L$,
\begin{equation}
P_{\rm corr} = \tilde{\phi_L}( E_{\rm iso}/E_{\rm pk}^{\alpha_E}/T_{r45,z}^{\alpha_T}/(1+z)^{\alpha_z}| \sigma_L ),
\end{equation}
without loss of generality.

In the case $\alpha_E=\alpha_T=\alpha_z=0$, $\tilde{\phi_L} = \tilde{\phi_L}(E_{\rm iso})$ is the luminosity function for $E_{\rm iso}$ smoothed by
a Gaussian kernel of width $\sigma_L$.  We think of $\tilde{\phi_L}$ as a luminosity function for the {\it effective}~luminosity $L$ (Equation \ref{eq:leff}).
With parameters $a$, $b$, $L_{\rm cut}$, and $\sigma_L$, we $\tilde \phi_L$ can take a wide variety of shapes: a sharply broken powerlaw ($\sigma_L\rightarrow 0$),
a Gaussian ($a_L$,$-b_L$ $\rightarrow \infty$), etc. 

\subsection{The Probability of Detection}

In the derivation of the observed event rate $r_{\rm obs} = \Theta(C-C_{\rm min}) r_{\rm true}$ in Section \ref{sec:redux}, we write $\Theta(C-C_{\rm min})$ as shorthand for the probability of detection of $C$ 
counts: it is zero if $C<C_{\rm min}$, one otherwise.  In practice, the count rate cutoff is not sharp.  More accurately,
for a GRB reaching BAT with a given effective count rate $C_{\rm eff} = C \sqrt{ f_p/T_{r45} }$ (Section \ref{sec:redux}), we determine the probability of detection as:
\begin{equation}
\Theta(C-C_{\rm min}) \Rightarrow P({\rm detect}|C_{\rm eff}) = {1 \over 2} + {1 \over 2}{\rm erf}\bigl( {\log{(C_{\rm eff}/C_{\rm eff,min})} \over \sigma_{C_{eff}}\sqrt{2}} \bigr),
\end{equation}
where ${\rm erf}$ is the error function.  This is the result of a convolution of $\Theta(C-C_{\rm min})$ in Equation \ref{eq:robs} with a log-Gaussian of width $\sigma_{C_{eff}}$.  The convolution allows the count rate cutoff to be smooth.

In principle, just as faint GRBs with low $C$ values are missing from the observed sample, some $C_{\rm eff,min}$
are also missing.  Our estimate of $0.24\pm 0.05$ above will then be biased \citep[see, e.g.,][]{pet96}.  It is necessary, therefore, to fit for
$C_{\rm eff,min}$ and $\sigma_{C_{\rm eff}}$.   In practice, we find that $C_{\rm eff,min}$ varies little (see, Table 1) and $\sigma_{C_{\rm eff}}$ can be fixed to 0.1 dex
without significantly affecting the other best-fit model parameters.

In the case of unknown partial coding fraction $p_f$, $P(\rm detect|C_{\rm eff})$ must be convolved with the $\log{(p_f)}$ distribution, which is well-described as an exponential with mean $-0.23$.
This situation arises in the calculation of the model normalization $A$ in equation \ref{eq:logl}.  Here, we also require a prescription for determining the expected
$C_{\rm eff}$ given intrinsic values of $z$, $E_{\rm iso}$, $E_{\rm pk}$, and $T_{r45}$.   This connection between input ($S_{\rm bol}$, $E_{\rm pk,obs}$) and
$C$ is found by fitting a smooth curve to the observed data for all GRBs in the sample.  We realized after the fact that these curves corresponds accurately
to one derived from
fixing a low energy \citep{band93} model spectral index of $-1.1$ and a high energy index of $-2.3$ for all GRBs. This is in keeping with the prior assumptions
(see Paper I) made in the spectral fitting.  The scatter between the observed $C$ and predicted $C$ given this spectrum (for fixed values of $S_{\rm bol}$ 
and $E_{\rm pk,obs}$) is less than 0.05 dex.  Therefore, to the extent that other sources of error dominate (e.g., that in $E_{\rm pk}$), it is acceptable in the
current study to assume all Swift GRBs have the same \citep{band93} model powerlaw spectral indices when normalizing the model in Equation \ref{eq:logl}.

\subsection{Model Fitting}

To fit the model $r_{\rm obs}(\vec \theta)$ to the observed data $D$ by finding the optimal parameters $\vec \theta$, we maximize the Poisson likelihood \citep[e.g.,][]{greg05}:
\begin{equation}
\mathcal{L}(D|\vec \theta) = A^N \prod_{i=0}^{N} r_i(\vec \theta) \exp\bigl[ -A\int d\vec\theta r(\vec \theta) \bigr],
\label{eq:logl}
\end{equation}
where $r_i$ is the model evaluated for the $i$th GRB, and $A$ is a normalization.  

Each GRB has an associated range of acceptable values for 
$E_{\rm pk}$, $E_{\rm iso}$, and $T_{r45}$ (Tables 3 \& 4),
and we average each $r_i$ over these ranges.
As discussed in Paper I, the spectral fitting returns highly correlated values for $E_{\rm pk}$ and $E_{\rm iso}$,
and this correlation is taken account via Monte Carlo integration as described in Paper I.
The integration of $r(\vec \theta)$ in the exponential in Equation \ref{eq:logl} for a given set of parameters $\vec \theta$ is independent of the data $D$ 
and is carried out numerically.
In the case of fitting GRBs without measured redshift (see Section \ref{sec:fitting}), we must also integrate over the
unknown $z$.  The redshift integration is carried out over the interval $z=(0.2,\infty)$ to match the selected sample range
(Section \ref{sec:redux}).  In practice, the above integrations can be carried out rapidly ($\Delta t\approx$ 100ms for a given $\vec \theta$ on a desktop PC),
and we utilize Markov Chain Monte Carlo techniques (Section \ref{sec:fitting}) to stochastically explore the allowed parameter space.  

Fitting is accomplished by maximizing Equation \ref{eq:logl} using Markov Chain Monte Carlo (MCMC) in python with PyMC\footnote{http://code.google.com/p/pymc}.
An MCMC method is preferred (relative, e.g., to direct maximization of Equation \ref{eq:logl})
due to the high dimensionality of the fitting problem ($17$ model parameters) and the natural tendency to find local maxima rather than the
global maximum.  Moreover, MCMC methods return joint confidence regions on all parameters directly, without requiring that these 
distributions be tabulated 
in a computationally intensive way apart from the fitting.  

Due to strong covariance among several model parameters (see below),
we utilize the ``Adaptive Metropolis'' algorithm in PyMC to efficiently
draw from the data posterior distribution.  This algorithm employs an estimate of the posterior covariance matrix --- based on the observed sample correlation
matrix calculated after every $10^3$ draws --- to randomly walk through the parameter space.  The presence of covariance in the Gaussian sampling 
distribution  allows for a high acceptance rate for the draws.  A scale factor is applied to the covariance matrix for sampling, and this is varied so that the
acceptance rate approaches a target of 30\%.  (Higher rates correspond to insufficient randomness in the sampling and failure to adequately explore the
parameter space.)  While the resulting chain is not strictly speaking Markov, it is ergodic.  We refer the interested reader to the PyMC user 
guide\footnote{http://pymc.googlecode.com/files/UserGuide2.0.pdf} for more details.

\begin{figure} 
\center{\includegraphics[width=4.0in]{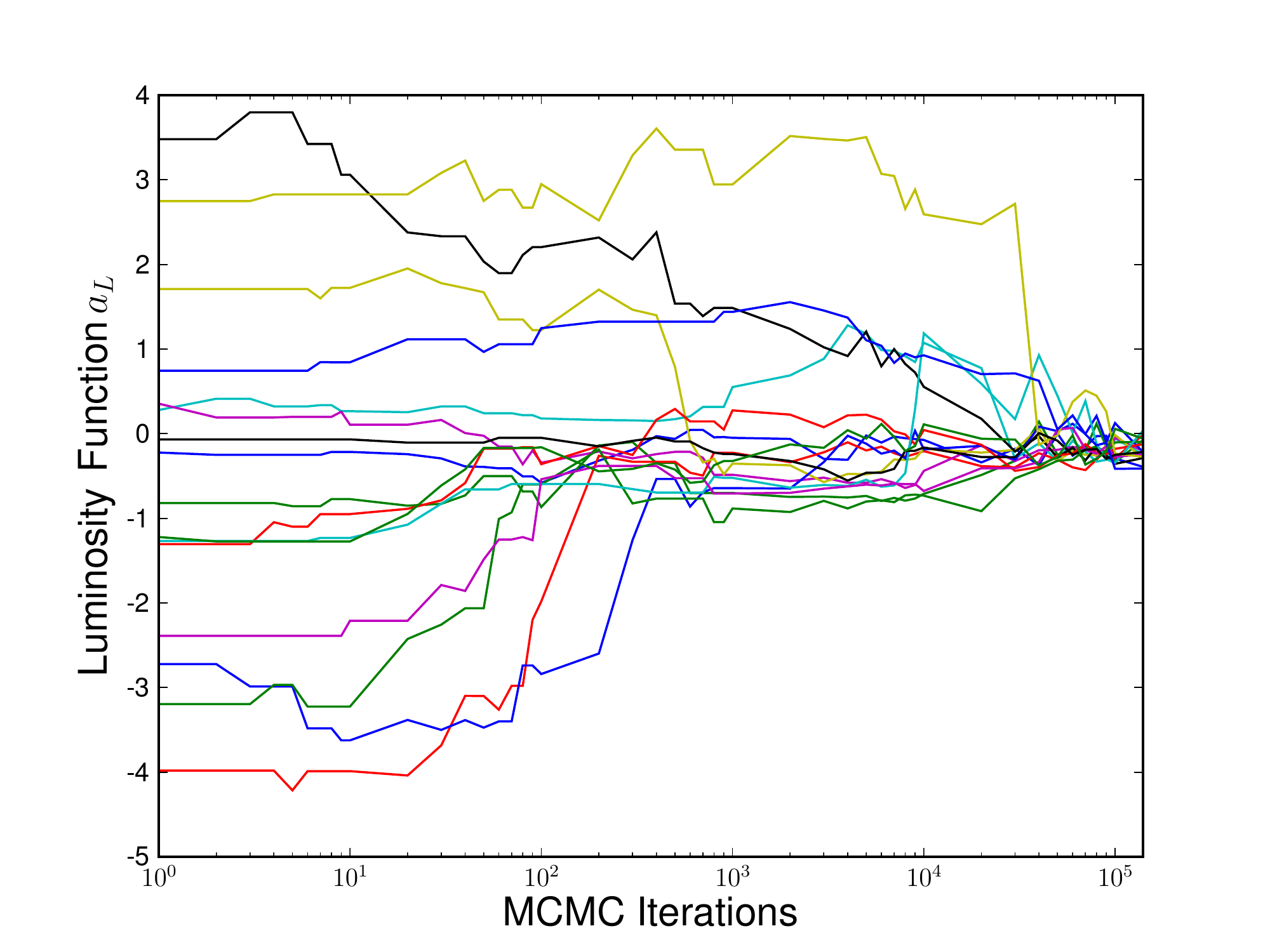}} 
\caption{\small 
Uniqueness of the solution as demonstrated by the convergence of 16 MCMC chains, started from random positions in the parameter space, shown here for the luminosity function index $a_L$.
Each chain converges to $a_L=-0.22$ and to the other parameters (not shown) in Table 1 (left) within $10^5$ iterations. 
A logarithmic sampling of each chain is plotted for ease of viewing.} 
\label{fig:mcmc} 
\end{figure}

We have initialized 16 chains randomly in the parameters space (Figure \ref{fig:mcmc}) and verified convergence to the values presented in Table 1.  Convergence typically
requires of order $2 \times 10^4$ iterations of each chain.  After convergence (``burn in'') we record one out of every ten successive iterations until
a total of $10^4$ samples are acquired.  This process of ``thinning'' the chain \citep[e.g.,][]{gelman04} mitigates against sample to sample correlation.
Best fit parameters and 90\% confidence error bars are reported in Table 1.  

\vfill
\eject

\begin{center}
\begin{scriptsize}
\begin{longtable}{l@{}c@{}c@{}cccccc}
\caption[]{BAT Trigger Statistics and Durations} \label{tab:time_table} \\

\hline \hline \\[2ex]
   GRB & Trigger Time & Burst Region & $T_{90}$ & $T_{50}$ & $T_{r45}$ & Rate$_p/$Cts & $dt_{S/N}$ & $S/N$ \\[0.5ex]
    & (UTC) & [s] & [s] & [s] & [s] & [$1/s$] & [s] & \\[0.5ex] \hline
   \\[1.8ex]
\endfirsthead

\multicolumn{8}{l}{{\tablename} \thetable{} -- Continued} \\[0.5ex]
  \hline \hline \\[2ex]
   GRB & Trigger Time & Burst Region & $T_{90}$ & $T_{50}$ & $T_{r45}$ & Rate$_p/$Cts & $dt_{S/N}$ & $S/N$ \\[0.5ex] \hline
  \\[1.8ex]
\endhead

  \\[1.8ex] \hline
  \multicolumn{8}{l}{{Continued on Next Page\ldots}} \\[0.5ex]
\endfoot

  \\[1.8ex] \hline \hline
\endlastfoot
070517 & 11:20:58.39 & $0.255\rightarrow 7.745$ & $6.2\pm0.4$ & $3.4\pm0.6$ & $1.5\pm0.2$ & $0.3\pm0.1$ & 6.86 & 8.7\\[0.5ex]
070518 & 14:26:21.20 & $-0.650\rightarrow 5.650$ & $5.3\pm0.3$ & $2.9\pm0.4$ & $1.2\pm0.1$ & $0.5\pm0.2$ & 5.94 & 10.1\\[0.5ex]
070520A & 13:05:10.03 & $25.565\rightarrow 61.205$ & $31\pm4$ & $12\pm4$ & $5.3\pm0.9$ & $0.06\pm0.05$ & 16.83 & 7.5\\[0.5ex]
070520B & 17:44:53.26 & $-5.175\rightarrow 31.185$ & $31\pm2$ & $15\pm2$ & $10\pm1$ & $0.05\pm0.02$ & 16.16 & 8.5\\[0.5ex]
070521 & 06:51:10.86 & $-17.245\rightarrow 594.555$ & $560\pm10$ & $16.8\pm0.6$ & $10.6\pm0.6$ & $0.074\pm0.005$ & 23.23 & 118.6\\[0.5ex]
070529 & 12:48:28.34 & $0.595\rightarrow 121.965$ & $112\pm4$ & $60\pm4$ & $23\pm2$ & $0.024\pm0.006$ & 121.37 & 13.5\\[0.5ex]
070531 & 02:10:17.52 & $-2.920\rightarrow 41.080$ & $37\pm4$ & $15\pm4$ & $6.6\pm0.7$ & $0.06\pm0.03$ & 20.20 & 14.4\\[0.5ex]
070610 & 20:52:26.14 & $-0.490\rightarrow 10.010$ & $8\pm1$ & $3.2\pm0.6$ & $1.6\pm0.2$ & $0.2\pm0.1$ & 5.58 & 12.7\\
\end{longtable}
{Notes: Detailed descriptions of the above quantities can be found in Paper I.}
\end{scriptsize}
\end{center}

\vfill
\eject

\begin{center}
\begin{scriptsize}
\begin{longtable}{@{}l@{}c@{}c@{}c@{}c@{}c@{}c@{}c@{}c@{}c}
\caption[]{BAT Spectral Fits} \label{tab:spec_table} \\

\hline \hline \\[2ex]
   GRB & -$\alpha$ & -$\beta$ & $E^{\rm freq.}_{\rm pk,obs}$ & $S_{15-350}$ & $\chi^2/\nu^{({\rm model})}$ & $E_{\rm pk,obs}$ & 
   $N_{\rm iso}$ or $n_{\rm bol}$ & $E_{\rm iso}$ or $S_{\rm bol}$ & $z^{({\rm ref})}$ \\[0.5ex]
    &  & & [keV] & [$10^{-6}{{\rm erg} \over {\rm cm}^2}$] &  & [keV] & [$10^{59}\gamma$ or $10^2{\gamma \over {\rm cm}^2}$] & 
    [$10^{52}$erg or $10^{-6}{{\rm erg} \over {\rm cm}^2}$] & \\[1.5ex] \hline
   \\[1.8ex]
\endfirsthead

\multicolumn{10}{l}{{\tablename} \thetable{} -- Continued} \\[0.5ex]
  \hline \hline \\[2ex]
   GRB & -$\alpha$ & -$\beta$ & $E^{\rm freq.}_{\rm pk,obs}$ & $S_{15-350~{\rm keV}}$ & $\chi^2/\nu^{({\rm model})}$ & $E_{\rm pk,obs}$ & 
   $N_{\rm iso}$ or $n_{\rm bol}$ & $E_{\rm iso}$ or $S_{\rm bol}$ & $z^{({\rm ref})}$ \\[0.5ex] \hline
  \\[1.8ex]
\endhead

  \\[1.8ex] \hline
  \multicolumn{10}{l}{{Continued on Next Page\ldots}} \\[0.5ex]
\endfoot

  \\[1.8ex] \hline \hline
\endlastfoot
070517 & $1.7\pm0.2$ & ... & $>$45 & $0.37\pm0.06$ & $47.9/56^{1}$ & $67^{+178}_{-35}$ & $0.13^{+0.12}_{-0.05}$ & $0.34^{+0.32}_{-0.07}$ & ...\\
070518 & ... & $2.1\pm0.3$ & $<$42 & $0.21\pm0.03$ & $66.1/56^{1}$ & $35\pm33$ & $0.3^{+1.9}_{-0.1}$ & $0.09^{+0.15}_{-0.01}$ & $1.16^{1}$\\
070520A & $0^{+1}_{-2}$ & ... & $70^{+180}_{-20}$ & $0.33^{+0.09}_{-0.06}$ & $52.8/55^{2}$ & $74^{+243}_{-35}$ & $0.14^{+0.14}_{-0.05}$ & $0.5^{+0.6}_{-0.1}$ & ...\\
070520B & $1.1\pm0.2$ & ... & $>$110 & $2.5\pm0.4$ & $54.8/56^{1}$ & $370^{+620}_{-220}$ & $0.32^{+0.13}_{-0.08}$ & $4^{+5}_{-2}$ & ...\\
070521 & $1.35\pm0.07$ & ... & $>$152 & $16.4\pm0.7$ & $52.7/56^{1}$ & $300^{+550}_{-160}$ & $3.5^{+0.9}_{-0.6}$ & $23^{+20}_{-8}$ & ...\\
070529 & $1.3\pm0.2$ & ... & $>$124 & $5.2\pm0.5$ & $44.3/56^{1}$ & $250^{+620}_{-130}$ & $5^{+4}_{-2}$ & $9^{+9}_{-3}$ & $2.4996^{2}$\\
070531 & $1.4\pm0.2$ & ... & $>$82 & $2.0\pm0.3$ & $57.8/56^{1}$ & $165^{+499}_{-73}$ & $0.37^{+0.26}_{-0.09}$ & $2.1^{+3.0}_{-0.6}$ & ...\\
070610 & $1.8\pm0.3$ & ... & ... & $0.41\pm0.08$ & $73.3/56^{1}$ & $74^{+424}_{-72}$ & $0.18^{+0.69}_{-0.05}$ & $0.43^{+0.67}_{-0.05}$ & ...\\
\end{longtable}
{Notes: 
The model in column 6 refers to a powerlaw (1), a powerlaw times exponential (2), a GRBM (3).  
If a redshift is known, it is given in the last table column and we present isotropic equivalent energy 
and photon fluences, $E_{\rm iso}$ and $N_{\rm iso}$, respectively, in columns 8 and 9.  Otherwise, we report approximate
bolometric fluences measured in the observer frame $1-10^4$ keV band in those columns. Additional details can be found in Paper I.}
{Redshift References:
$^{1}$\citet{zref81},
$^{2}$\citet{zref82},
$^{3}$\citet{zref83},
$^{4}$\citet{zref84},
$^{5}$\citet{zref85},
$^{6}$\citet{zref86},
$^{7}$\citet{zref87},
$^{8}$\citet{zref88},
$^{9}$\citet{zref89},
$^{10}$\citet{zref90},
$^{11}$\citet{zref91},
$^{12}$\citet{zref92},
$^{13}$\citet{zref93},
$^{14}$\citet{zref94},
$^{15}$\citet{zref95},
$^{16}$\citet{zref96},
$^{17}$\citet{zref97},
$^{18}$\citet{zref98},
$^{19}$\citet{zref99},
$^{20}$\citet{zref100},
$^{21}$\citet{zref101},
$^{22}$\citet{zref102},
$^{23}$\citet{zref103},
$^{24}$\citet{zref104},
$^{25}$\citet{zref105},
$^{26}$\citet{zref106},
$^{27}$\citet{zref107},
$^{28}$\citet{zref108},
$^{29}$\citet{zref109},
$^{30}$\citet{zref110},
$^{31}$\citet{zref111},
$^{32}$\citet{zref112},
$^{33}$\citet{zref113},
$^{34}$\citet{zref114},
$^{35}$\citet{zref115},
$^{36}$\citet{zref116},
$^{37}$\citet{zref117},
$^{38}$\citet{zref118},
$^{39}$\citet{zref119},
$^{40}$\citet{zref120},
$^{41}$\citet{zref121},
$^{42}$\citet{zref122},
$^{43}$\citet{zref123},
$^{44}$\citet{zref124},
$^{45}$\citet{zref125},
$^{46}$\citet{zref126},
$^{47}$\citet{zref127},
$^{48}$\citet{zref128},
$^{49}$\citet{zref129},
$^{50}$\citet{zref130},
$^{51}$\citet{zref131},
$^{52}$\citet{zref132},
$^{53}$\citet{zref133},
$^{54}$\citet{zref134},
$^{55}$\citet{zref135},
$^{56}$\citet{zref136},
$^{57}$\citet{zref137},
$^{58}$\citet{zref138},
$^{59}$\citet{zref139},
$^{60}$\citet{zref140},
$^{61}$\citet{zref141},
$^{62}$\citet{zref142},
$^{63}$\citet{zref143},
$^{64}$\citet{zref144},
$^{65}$\citet{zref145},
$^{66}$\citet{zref146},
$^{67}$\citet{zref147},
}
\end{scriptsize}
\end{center}

\end{document}